\documentclass[journal]{IEEEtran}
 \usepackage{amsmath,amssymb}
 \usepackage{subfigure}
 \usepackage{graphicx,graphics,color,psfrag}
 \usepackage{cite,balance}
 \usepackage{algorithm}
 \usepackage{accents}
 \usepackage{amsthm}
 \usepackage{bm}
 \usepackage{url}
 \usepackage{algorithmic}
 \usepackage[english]{babel}
 \usepackage{multirow}
 \usepackage{enumerate}
 \usepackage{cases}
 \usepackage{stfloats}
 \usepackage{dsfont}
 \usepackage{color,soul}
 \usepackage{amsfonts}
  \usepackage{tcolorbox}
\usepackage[utf8]{inputenc}
 \usepackage{cite,graphicx,amsmath,amssymb}
 \usepackage{subfigure}
 \usepackage{fancyhdr}
 \usepackage{hhline}
 \usepackage{graphicx,graphics}
 \usepackage{array,color}
 \usepackage{amsmath}
 \usepackage{amsthm}

\newtheorem{proposition}{Proposition}
\newtheorem{remark}{Remark}
\newtheorem{theorem}{Theorem}
\newtheorem{lemma}{Lemma}

\newtheorem{assumption}{Assumption}
\newtheorem{corollary}{Corollary}

\include{header}
\IEEEoverridecommandlockouts

\begin{document}

\title{Optimized Power Control Design for Over-the-Air Federated Edge Learning}
\author{Xiaowen Cao,  \emph{Student Member, IEEE,} Guangxu Zhu,  \emph{Member, IEEE,} Jie Xu,  \emph{Member, IEEE,} Zhiqin Wang, and Shuguang~Cui, \emph{Fellow, IEEE}\\
\thanks{Part of this paper has been presented at the IEEE International Conference on Communications (ICC), Montreal, Canada, Jun. 14-23, 2021 \cite{Cao_ICC21}. 
This work was supported by the Key Area R\&D Program of Guangdong Province with grant No. 2018B030338001, the National Key R\&D Program of China with grant No. 2018YFB1800800, Shenzhen Outstanding Talents Training Fund, Guangdong Research Project No. 2017ZT07X152, the National Natural Science Foundation of China under grants U2001208, 61871137, and 62001310, and the Science and Technology Program of Guangdong Province under grant 2021A0505030002. Corresponding authors: Guangxu Zhu and Zhiqin Wang. 
}
\thanks{X. Cao is with the School of Information Engineering, Guangdong University of Technology, Guangzhou, China, and the Future Network of Intelligence Institute (FNii), The Chinese University of Hong Kong (Shenzhen), Shenzhen, China (e-mail: caoxwen@outlook.com). }
\thanks{
G. Zhu is with Shenzhen Research Institute of Big Data, Shenzhen, China  (e-mail: gxzhu@sribd.cn).  }
\thanks{
J. Xu is with the School of Science and Engineering (SSE) and the FNii, The Chinese University of Hong Kong (Shenzhen), Shenzhen, China (e-mail: xujie@cuhk.edu.cn). }
\thanks{Z. Wang is with China Academy of Information and Communications Technology, Beijing, China (e-mail: zhiqin.wang@caict.ac.cn).}
\thanks{S. Cui is with the SSE and the FNii, The Chinese University of Hong Kong (Shenzhen), Shenzhen, China, and Shenzhen Research Institute of Big Data, Shenzhen, China. (e-mail: shuguangcui@cuhk.edu.cn).  }
}

\markboth{}{}
\maketitle

\setlength\abovedisplayskip{2pt}
\setlength\belowdisplayskip{2pt}

\begin{abstract}%\vspace{-0.3cm}
\emph{Over-the-air federated edge learning} (Air-FEEL) has emerged as a communication-efficient solution to enable distributed machine learning over edge devices by using their data locally to preserve the privacy.
By exploiting the waveform superposition property of wireless channels, Air-FEEL allows the ``one-shot'' over-the-air aggregation of gradient-updates to enhance the communication efficiency, but at the cost of a compromised learning performance due to the aggregation errors caused by channel fading and noise. This paper investigates the transmission power control to combat against such aggregation errors in Air-FEEL. Different from conventional power control designs (e.g., to minimize the individual \emph{mean squared error} (MSE) of the over-the-air aggregation at each round), we consider a new power control design aiming at directly maximizing the convergence speed. Towards this end, we first analyze the convergence behavior of Air-FEEL (in terms of the optimality gap) subject to aggregation errors at different communication rounds. It is revealed that if the aggregation estimates are unbiased, then the training algorithm would converge exactly to the optimal point with mild conditions; while if they are biased, then the algorithm would converge with an error floor determined by the accumulated estimate bias over communication rounds. Next, building upon the convergence results, we optimize the power control to directly minimize the derived optimality gaps under the cases without and with unbiased aggregation constraints, subject to a set of average and maximum power constraints at individual edge devices. 
We transform both problems into convex forms, and obtain their structured optimal solutions, both appearing in a form of regularized channel inversion, by using the Lagrangian duality method. Finally, numerical results show that the proposed power control policies achieve significantly faster convergence for Air-FEEL, as compared with benchmark policies with fixed power transmission or conventional MSE minimization.
\end{abstract}
%}

\begin{IEEEkeywords}%\vspace{-0.4cm}
Federated learning, over-the-air computation, stochastic gradient descent, power control, edge intelligence.
\end{IEEEkeywords}

%%\vspace{-0.2cm}
\section{Introduction}\label{sec:intro}
In the pursuit of ubiquitous brain-inspired intelligence envisioned in the future 6G networks \cite{KLetaif19_6G}, recent years have witnessed the spreading of {\it artificial intelligence} (AI) algorithms from the cloud to the network edge, resulting in an active area called edge intelligence \cite{Survey_FEEl,Debbah19}.
%Edge intelligence, referring to deploying the  {\it artificial intelligence} (AI) algorithms at the network edge (e.g., base stations and mobile devices) \cite{Survey_FEEl,Debbah19}, has been recognized as one of the key technologies in 6G networks  to realize the ubiquitous intelligence vision.
The core research issue therein is to allow low-latency and privacy-aware access to rich mobile data for intelligence distillation.
To this end, the {\it federated edge learning} (FEEL) framework has been proposed recently, which distributes the AI-model training task over edge devices by using their data locally to preserve the privacy \cite{Tran19_Infocom,Konecny2016aa_FL,Chen20FL,JParK21Proc,Mo2020aa,JRen20}.
Essentially, the FEEL framework corresponds to the implementation of {\it distributed gradient descent} over wireless networks. 
Such a training process is to find optimized AI-model parameters by minimizing a properly designed loss function in an iterative manner.
Specifically, at each iteration or communication round, the edge server first broadcasts the global AI-model parameters to edge devices, such that all edge devices can synchronize their local models; 
next, the edge devices compute their respective local gradient updates using the local data and then upload them to the edge server for further aggregation to update the global model.
Although the uploading of high-volume raw data is avoided, the gradient aggregation process in FEEL may still suffer from a communication bottleneck due to the high dimensionality of each gradient update, especially when the number of edge devices sharing the same wireless medium becomes large. 
To tackle this issue, one promising solution called {\it over-the-air} FEEL  (Air-FEEL) has been proposed (see, e.g., \cite{Zhu2021ComMag}), which exploits the {\it over-the-air computation} (AirComp) technique for ``one-shot" aggregation by allowing multiple devices' concurrent update transmission. In such a way, the communication and computation are integrated in a joint design by exploiting  the superposition property of a {\it multiple access channel} (MAC)  \cite{nomo_function_Nazer,Gastpar08}.
%}

%\subsection{Over-the-air Computation}%\vspace{-0.1cm}

The idea of AirComp was first proposed in \cite{nomo_function_Nazer} for data aggregation in sensor networks, which harnessed the ``interference" via structured codes to help functional computation over a MAC. The subsequent work \cite{Gastpar08} showed that for Gaussian {\it independent and identically distributed} (i.i.d.) data sources, the uncoded (analog) transmission is optimal to minimize the distortion in AirComp. 
Building on the information-theoretic studies, the analog AirComp implementation has attracted growing research interests (see, e.g., \cite{Abari15,Cao_PowerTWC,WLiu20TWC,Cao_2020aa,GX18IoT,LChen18IoT,XZhai2020Ar,ZD18TSP}). For instance, the synchronization issue in AirComp was addressed in \cite{Abari15} via a shared clock broadcasting from edge server to devices; the optimal power control policies for AirComp over fading channels were derived in \cite{Cao_PowerTWC,WLiu20TWC} to minimize the average computation distortion; and the cooperative interference management for multi-cell AirComp networks was developed in \cite{Cao_2020aa}.
Furthermore, {\it multiple-input-multiple-output} (MIMO) spatial multiplexing was exploited in AirComp to enable vector-valued functional computation targeting multi-modal sensing \cite{GX18IoT,LChen18IoT} and to enhance the computational accuracy \cite{XZhai2020Ar,ZD18TSP}. 
%A blind MIMO AirComp was proposed in \cite{} without requiring {\it channel state information} (CSI). 
%In addition, UAV-aided AirComp system was studied in \cite{Fu2021Ar}, while an wireless powered AirComp in an {\it intelligent reflecting surface} (IRS) system was investigated in \cite{ZW21IoT} 
%The channel feedback overhead in MIMO AirComp was then exempted in \cite{ZD18TSP}, by solving a bilinear estimation problem that can recover both the channel information and the desired functions simultaneously from a set of noisy received aggregated signals. 

%\subsection{AirComp for Federated Edge Learning}%\vspace{-0.1cm}

Recently, AirComp found its merits in the new context of FEEL, namely the Air-FEEL, to enable the communication-efficient gradient aggregation at each communication round \cite{TSery2020TSP}. 
%In the literature, there have been several 
Prior works studied the Air-FEEL system from different perspectives such as devices scheduling \cite{GZhu2020TWC,YSun2021Ar,SXia2020Ar,XFan2021Ar04}, beamforming design \cite{ KYang2020TWC,SWang2021Ar}, update compression \cite{Amiri2020TSP,MAmiri2020TWC,XFan2021Ar}, and hyper-parameters (such as learning rates) optimization \cite{HG21IoT,Xu2021Ar,JZhang2021Ar}.
For instance, a broadband Air-FEEL solution was proposed in \cite{GZhu2020TWC}, where a set of communication-learning tradeoffs were derived to guide the device scheduling. 
Along this vein, the authors in \cite{YSun2021Ar} proposed an energy-aware device scheduling strategy to minimize the expected improvement of loss function value at each communication round, while the authors in \cite{SXia2020Ar} proposed a threshold-based device selection scheme to achieve reliable model aggregation.
Then, for multi-antenna Air-FEEL systems, a joint design of device scheduling and receive beamforming was presented in \cite{KYang2020TWC}, and a unit-modulus analog receive beamforming design was proposed in \cite{SWang2021Ar}.
As for update compression, a source-coding algorithm exploiting gradient sparsification was proposed in \cite{Amiri2020TSP,MAmiri2020TWC}, and a compressive-sensing based gradient aggregation approach was developed in \cite{XFan2021Ar} to further improve the communication efficiency.
Lately, Air-FEEL based on digital modulation was proposed in \cite{GZhu2020Ar} and further extended in \cite{RJiang2020Ar}, which features one-bit quantization and modulation at the edge devices and majority-vote based decoding at the edge server.
%while a 1-bit compressive sensing based Air-FEEL system was considered in \cite{XFan2021Ar} to minimize the expected convergence rate through the joint optimization problem of computation.
%The authors in \cite{TSery2020TSP} proved that the gradient-based multiple access algorithm can achieve the same convergence rate as the centralized gradient descent algorithm in large-scale networks. 
%Subsequently, the gradient statistics aware power control was investigated in \cite{NZhang2020Ar} to further enhance the performance of Air-FEEL. 
%Additionally, the learning performance of Air-FEEL can be further enhanced by gradient-aware power control \cite{NZhang2020Ar},  %joint device selection and power control design \cite{Fan2021Ar}, 
%and compressive-sensing based gradient aggregation approach \cite{XFan2021Ar}.
Besides improving the communication efficiency, the Air-FEEL has also been exploited for enhancing the data privacy, in which individual updates are not accessible by the centralized edge server, thus eliminating the risk of potential model inversion attack \cite{DLiu2020Ar,Koda2020Ar,ASonee2020Ar,MSeif2020Ar}.

Generally speaking, the employment of AirComp introduces an essential design tradeoff in Air-FEEL between the the enhanced communication efficiency (via the over-the-air aggregation) and the degraded learning performance (due to the aggregation error caused by the channel fading and noise perturbation). 
Due to such a tradeoff, how to analytically characterize the training performance (e.g., in terms of accuracy and latency) is a challenging task has not been investigated in the literature yet.
Also notice that the aggregation distortion in different communication rounds may have distinct impacts on the learning performance \cite{CShen2021Ar}, thus making the above tradeoff even more complicated. 
%However, by taking these effects into consideration, how to analytically characterize the training performance (e.g., in terms of accuracy and latency) has not been investigated in the literature yet. 
To deal with these issues,  it is crucial to properly control the transmission power at different edge devices over different communication rounds. 
In the literature, there have been prior works on Air-FEEL \cite{GZhu2020TWC,YSun2021Ar,SXia2020Ar,Amiri2020TSP,KYang2020TWC,DLiu2020Ar,Xu2021Ar,XFan2021Ar04} that considered simplified channel inversion (or its variants)  to align the channel gains among different devices, which, however, may lead to amplified noise at receiver and thus is highly suboptimal especially in deep fading scenarios. 
 Some other works \cite{NZhang2020Ar,JZhang2021Ar} designed the power control with the objective of minimizing the individual aggregation distortion (e.g., {\it mean squared error} (MSE)) at each communication round, which, however, may not perform well as the aggregation distortion in different communication rounds may have distinct impact on the learning performance \cite{CShen2021Ar}. 
Therefore, how to obtain the  
analytic learning performance of Air-FEEL in terms of the power control variables, and accordingly optimize the power control decisions for optimizing the learning performance still remains unknown. 
This thus motivates the current work.

%\subsection{Main Contributions}%\vspace{-0.1cm}

This paper studies an Air-FEEL system consisting of multiple edge devices and one edge server. By considering smooth learning models satisfying the Polyak-{\L}ojasiewicz inequality, we establish an elegant learning performance metric, namely the optimality gap, linking with the aggregation errors over communication rounds.
Accordingly, we  propose optimized power control policies for directly minimizing the optimality gap.
The main contributions are summarized as follows.
\begin{itemize}
	\item { \bf Convergence analysis:}
	First, we analyze the optimality gap of the loss function over different communication rounds, which characterizes the impact of gradient aggregation errors (i.e., the bias and MSE of the gradient aggregation estimates) on the convergence performance of the Air-FEEL algorithm. It is revealed that if the aggregation errors are unbiased, the optimality gap will diminish to zero with sufficiently many communication rounds and properly chosen step sizes; while if the aggregation errors are biased, the optimality gap would reach to an error floor whose height is equal to the accumulated estimate bias over communication rounds. 
	It is also shown that within a finite number of communication rounds, the aggregation errors at later rounds  (with higher weights) contribute more to the optimality gap than those at earlier rounds.
	\item {  \bf Power control optimization:}
	Next, building on the convergence results, we formulate new power control optimization problems for Air-FEEL under the cases without and with unbiased aggregation constraints, respectively, with the objective of minimizing the optimality gap, subject to a set of average and maximum power constraints at individual edge devices. 
	Fortunately, both power control problems can be transformed into convex forms, which can thus be optimally solved by the  Lagrangian duality method. The optimized power control solutions establish a regularized channel inversion structure, where the regularization term at each edge device is related to all other devices' average power budgets for the case without unbiased aggregation constraints, and is only related to the own device's individual average power budget for the case with unbiased aggregation constraints. 
   \item  {\bf Performance evaluation:} Finally, we conduct extensive simulations to evaluate the performance of the optimized power control for Air-FEEL by considering the ridge regression with synthetic dataset, and handwritten digit recognition using MNIST dataset with a {\it convolution neural network} (CNN). 
   It is shown that the proposed power control policies in the cases without and with unbiased aggregation constraints achieve significantly faster convergence rate (or lower optimality gap), than the benchmarking fixed power transmission and conventional MSE minimization schemes, as the proposed policies can better handle the aggregation errors over rounds based on their contributions to the optimality gap. It is also shown that the the case without unbiased aggregation constraints can achieve lower optimality gap than that with biased aggregation constaints when some mild conditions are met. This validates our analysis that the Air-FEEL can always converge to the optimal point with unbiased gradient aggregation.
\end{itemize}

{\it Notations:}  Bold lowercase letters refer to column vectors.
$\mathbb{E}(\cdot)$ denotes the expectation operation; the superscript $T$ represents the transpose operation; $\nabla$ is the gradient operator, and $(x)^+\triangleq\max\{0,x\}$.
For a set $\mathcal{A}$, $|\mathcal{A}|$ denotes its cardinality.
$\|\bm a\|$ denotes the Euclidean norm of vector $\bm a$. $\bf I$ denotes the identity matrix.
%$|\bm A|$ denotes the determinant of a squared matrix $\bm A$.
 For ease of reference, the main notations used in this paper are listed in Table \ref{t}.

\begin{table}[h]%\vspace{-0.5cm}
\caption{List of Main Notations}\label{t}%\vspace{-1cm}
\begin{center}
\begin{tabular}{|m{1.3cm}<{\centering}| p{6.6cm}|}
\hline
Symbol & Description\\ \hline
$\mathcal{K}$ & Set of edge devices with $\mathcal{K}\triangleq  \{1,..., K\}$ \\ \hline
%$N$ & Number of needed communication rounds\\ \hline
$\mathcal{N}$ & Set of communication rounds with $ \mathcal{N}\triangleq\{ 1,\cdots,N\}$\\ \hline
${\bf w}\in\mathbb{R}^q$ &  Parameter vector of learning model with size $q$ \\ \hline
${\bf w}^{\star}$ &  Optimal parameter vector \\ \hline
${\mathcal D}_k$ &  Local dataset at edge device $k\in\mathcal{K}$ with cardinality $D_k$\\ \hline
$({\bf x}_i,\tau_i)$& The $i$-th sample ${\bf x}_i$ in dataset with ground-true label $\tau_i$\\ \hline
  $f({\bf w},{\bf x}_i,\tau_i)$ & Sample-wise loss function for quantifying the prediction error of the learning model $\bf w$ on ${\bf x}_i$ in terms of $\tau_i$\\ \hline
$F_k({\bf w})$ &  Local loss function of the learning model vector $\bf w$ on ${\mathcal D}_k$ at edge device $k\in\mathcal{K}$ \\ \hline
%$R({\bf w})$ & Regularization function\\ \hline
$F({\bf w})$ & Global loss function at the parameter model $\bf w$  \\ \hline
 ${\bf g}_{k}^{(n)}$ & Local gradient estimate in edge device $k$ at communication round $n$\\ \hline
 $\bar{\bf g}^{(n)} $ & Global gradient estimation at communication round $n$\\ \hline
$\hat{\bf g}^{(n)}$ &  Global gradient received at the edge server through over-the-air aggregation at communication round $n$ \\ \hline
$\eta^{(n)}$ & Learning rate at communication round $n$ \\ \hline
${\bm \varepsilon}^{(n)}$& Aggregation error at each communication round $n$ \\ \hline
 $\nabla F({\bf w})$  &  Ground-truth gradient of the loss function evaluated at point ${\bf w} \in\mathbb{R}^q$\\ \hline
 $F^{\star}$ & Optimal loss function value that is equal to $F({\bf w}^{\star})$\\ \hline
$\hat h_k^{(n)}$ & Complex channel coefficient from edge device $ k\in{\mathcal K}$ to the edge server at communication round $n$ \\ \hline
${\bf z}^{(n)}\in\mathbb{R}^q$ & AWGN with ${\bf z}^{(n)}\sim{\mathcal CN}(0,\sigma_z^2\bf I)$\\ \hline
 $p^{(n)}_{k}$ &  Power scaling factor in edge device $k\in\mathcal{K}$ at communication round $n$ \\ \hline
$ {\bf y}^{(n)}$ &  Received signal in edge server at communication round $n$ \\ \hline
%${\Phi}(\{p_k^{(n)}\})$& Upper bound of the optimality gap in the biased aggregation case \\ \hline
%$\gamma^{(n)}$ & Scaling factor among edge devices at communication round $n$ \\ \hline
\end{tabular}
\end{center}%\vspace{-1.2cm}
\end{table}

\section{System Model}\label{sec:system}

\begin{figure}
\centering
 \setlength{\abovecaptionskip}{-4mm}
\setlength{\belowcaptionskip}{-4mm}
    \includegraphics[width=3.7in]{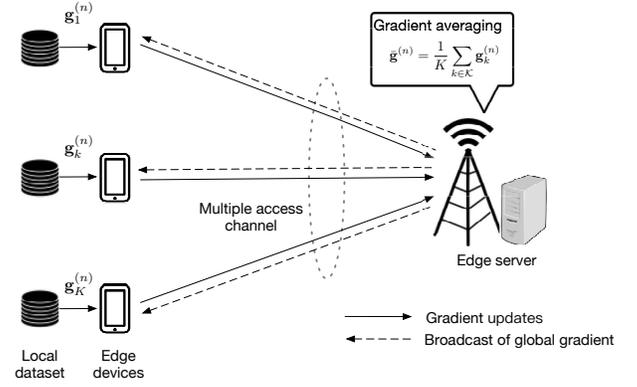}
\caption{Illustration of over-the-air federated edge learning (Air-FEEL). } \label{fig:model}
%-\vspace{-0.4cm}
\end{figure}

We consider an Air-FEEL system consisting of an edge server and $K$ edge devices, as shown in Fig.~\ref{fig:model}. With the coordination of the edge server, the edge devices cooperatively train a shared machine learning model via the over-the-air gradient aggregation as elaborated in the sequel.

\subsection{Learning Model}
We assume that the learning model is represented by the parameter vector ${\bf w}\in\mathbb{R}^q$ with ${\bf w}=[w_1,\cdots,w_q]^T$ and $q$ denoting the learning model size. 
%Here, the superscript $T$ denotes the transpose operation.
Let ${\mathcal D}_k$ denote the local dataset at edge device $k$, in which the $i$-th sample and its ground-true label are denoted by ${\bf x}_i$ and $\tau_i$, respectively. Define $f({\bf w},{\bf x}_i,\tau_i)$ as the sample-wise loss function quantifying the prediction error of the learning model $\bf w$ on sample ${\bf x}_i$ {\it with respect to} (w.r.t.) its ground-true label $\tau_i$.
Then the local loss function of the learning model vector $\bf w$ on ${\mathcal D}_k$ is
\begin{align}\label{LocalLossFunction}
F_k({\bf w})=\frac{1}{|{\mathcal D}_k|} \sum \limits_{({\bf x}_i,\tau_i)\in{\mathcal D}_k} f({\bf w},{\bf x}_i,\tau_i).
\end{align}
%where $R({\bf w})$ denotes the strongly convex regularization function whose strength is controlled by a hyperparameter $\rho\geq 0$.
For notational convenience, we denote $f({\bf w},{\bf x}_i,\tau_i)$ as $f_i({\bf w})$ and assume that the sizes of local datasets at different edge devices are uniform, i.e., $D \triangleq D_k=|{\mathcal D}_k|,\forall k\in\mathcal K$.
Then, the global loss function on all the distributed datasets  ${\mathcal D}_{\rm tot}=\cup_{k\in\mathcal K} {\mathcal D}_k$ evaluated on parameter vector $\bf w$ is given by
\begin{align}\label{GlobalLossFunction}
F({\bf w})=\frac{1}{ D_{\rm tot}}\sum\limits_{k\in\mathcal K} D_k F_k({\bf w})=\frac{1}{ K}\sum\limits_{k\in\mathcal K} F_k({\bf w}),
\end{align}
where $D_{\rm tot}=|{\mathcal D}_{\rm tot} |=KD$.

The objective of the training process is to find a desired parameter vector $\bf w$ for minimizing the global loss function $F({\bf w})$ in \eqref{GlobalLossFunction}, i.e., 
 \begin{align}\label{OptimalParameter}
 {\bf w}^{\star}=\arg \min_ {\bf w} F({\bf w}).
\end{align}
Instead of uploading all the local data to the edge server for centralized training, we consider the Air-FEEL, in which the learning process in \eqref{OptimalParameter} is implemented iteratively in a distributed manner using the {\it federated stochastic gradient descent} (FedSGD) algorithm\footnote{Besides the FedSGD, the {\it federated averaging} (FedAvg) is an alternative method for Air-FEEL, which can be implemented with multiple local updates at edge devices together with the model averaging at the edge server. This paper considers the FedSGD (instead of  FedAvg), mainly for the purpose of facilitating the convergence analysis and gaining insightful results. Furthermore,  the FedSGD enjoys the advantages of being more robust to the non-i.i.d. data \cite{Hou2021_Ar}. Nevertheless, our design and analysis principles in this paper can be extended to the case with FedAvg, which is left for future work.} \cite{FedAvg} as detailed in the following.
%as illustrated in Fig.~\ref{fig:model}.

%{\color{red} The gradient descent method should be improved to SGD method}

Consider a particular iteration or communication round $n$, with the learning model before updating being denoted by ${\bf w} ^{(n)}$. In this round, each edge device $k\in \mathcal K$ computes the local gradient estimate of the loss function as ${\bf g}_{k}^{(n)}$, based on a randomly sampled mini-batch from the local dataset. We denote the set of mini-batch data used by the edge device $k$ at round $n$ as $ \tilde{\mathcal D}_k^{(n)}$ and its size  $m_b = | \tilde{\mathcal D}_k^{(n)}|, \forall k\in \mathcal K$.
%Let $ \tilde{\mathcal D}_k \subset {\mathcal D}_k $ denote the randomly selected dataset by device $k$, where $m_b = | \tilde{\mathcal D}_k|, \forall k\in \mathcal K$, represents the mini-batch size. 
Then we have 
\begin{align}\label{sys_LocalGradient}
{\bf g}_{k}^{(n)}=	\frac{1}{m_b} \sum \limits_{({\bf x}_i,\tau_i)\in\tilde{\mathcal D}_k} \nabla f_i\left({\bf w}^{(n)}\right).
\end{align}
%Notice that, $m_b=D= |{\mathcal D}_k |$ corresponds to the case when  the whole local datasets at devices are used to compute the local gradient estimates.
Next, the edge devices upload their local gradients to the edge server for aggregation.
If the aggregation is error-free, then the global gradient estimate can be obtained as an average of local gradient estimates from all different edge devices, i.e.,\footnote{ Although we consider the same data size $D$ at  different edge devices, our proposed Air-FEEL can be easily extended to the case when they have different data sizes, i.e., $D_k$'s are different. In this case, we only need to revise the global gradient estimate in \eqref{sys_GlobalGradient} as a weighted-average of the local ones, i.e., $\bar{\bf g}^{(n)}=\sum\limits_{k\in\mathcal K}\frac{D_k}{D_{\rm tot}}{\bf g}_{k}^{(n)}$. Via AirComp, the desired weighted aggregation of the local gradient estimate can be  easily attained by adding an additional pre-processing $\psi(\cdot)$ on the transmitted signal $s_k$ with $\psi(s_k)=\sum\limits_{k\in\mathcal K}\frac{D_k}{D_{\rm tot}}s_k$.}
\begin{align}\label{sys_GlobalGradient}
\bar{\bf g}^{(n)}=	\frac{1}{K}\sum\limits_{k\in\mathcal K} {\bf g}_{k}^{(n)}.
\end{align}
Then, the edge server broadcasts the obtained global gradient estimate $\bar{\bf g}^{(n)} $ to the edge devices, based on which different edge devices can synchronously update their own learning model via
\begin{align}\label{sys_ModelUpdate}
{\bf w}^{(n+1)}={\bf w}^{(n)}-\eta^{(n)}\cdot \bar{\bf g}^{(n)},
\end{align}
where $\eta^{(n)}$ is the learning rate at communication round $n$. The above procedure continues until the convergence criteria is met or the maximum number of communication rounds is achieved.

%{\color{blue}

Notice that this paper considers the over-the-air aggregation approach to achieve fast gradient aggregation, based on which the received aggregated gradient at the edge server in \eqref{sys_GlobalGradient} may be erroneous due to perturbation caused by the channel fading and noise. This issue will be elaborated in Section~\ref{CommunicationModel}. 

%}

 \subsection{Basic Assumptions on Learning Model}
 To facilitate the convergence analysis, we make several assumptions on the loss functions and gradient estimates, which are commonly made in the literature \cite{DLiu2020Ar,JZhang2021Ar,SXia2020Ar,SWang2021Ar,Michael2012,Li2019_Noniid}.

 \begin{assumption}[Smoothness]\label{Assump_Smooth}\emph{
Let $\nabla F({\bf w})$ denote the gradient of the loss function evaluated at point ${\bf w} \in\mathbb{R}^q$. Then there exists a non-negative constant vector  ${\bf L}\in\mathbb{R}^q$ with ${\bf L}=[L_1,\cdots,L_q]^T$, such that
\begin{align*}
& F({\bf w})\!-\!\left[ F({\bf w}^{\prime})\! +\! \nabla F({\bf w})^T (\!{\bf w}\!\!-\! {\bf w}^{\prime})\right]\\
&~~~~~~~~~~~~~\le \frac{1}{2}\sum_{i=1}^{q} L_i({{ w}_i-{w}^{\prime}_i})^2, \forall {\bf w}, {\bf w}^{\prime} \in\mathbb{R}^q.
%\sum_{i=1}^{q}\! \!L_i({{ w}_i-\!{w}^{\prime}_i})^2\!, \forall {\bf w}, {\bf w}^{\prime},
\end{align*}}
\end{assumption}
 Assumption \ref{Assump_Smooth} guarantees that the gradient of  the loss function would not change arbitrarily quickly w.r.t. the parameter vector. Note that such an assumption is essential for convergence analysis of gradient decent methods to provide a good indicator for how far to decrease to the minimum loss.

\begin{assumption}[Polyak-{\L}ojasiewicz inequality]\label{Assump_PL}\emph{
Let $F^{\star}$ denote the optimal loss function value to problem \eqref{OptimalParameter}. There exists a constant $\delta\ge 0$ such that the global loss function $F({\bf w})$ satisfies the following Polyak-{\L}ojasiewicz condition:
\begin{align}\label{Ineq_PL}
	\| \nabla F({\bf w}) \|^2 \ge 2\delta(F({\bf w})-F^{\star}).
\end{align}
%where $\delta\ge 0$ is a constant.
}
\end{assumption}
Notice that Assumption \ref{Assump_PL} is more general than the standard assumption of strong convexity \cite{Karimi2016}. The inequality in \eqref{Ineq_PL} simply requires that the gradient grows faster than a quadratic function when away from the optimal function value and implies that every stationary point is a global minimum.
 Typical loss functions satisfying Assumptions \ref{Assump_Smooth}  and \ref{Assump_PL} include logistic regression, linear regression, and least squares.

 \begin{assumption}[Variance bound]\label{Assum_VarianceBound}\emph{
The local gradient estimates $\{{\bf g}_k\}$, defined in \eqref{sys_LocalGradient}, where the index $n$ is omitted for simplicity, are assumed to be independent and unbiased estimates of the batch gradient $\nabla F({\bf w})$ with coordinate bounded variance, i.e.,
\begin{align}
	&\mathbb{E}[{\bf g}_k]=\nabla F({\bf w}), \forall k\in\mathcal K,\\
	&\mathbb{E}[ ({g}_{k,i}-\nabla F({w_i}) )^2]\le \frac{\sigma_i^2}{m_b}, \forall k\in\mathcal K, \forall i,
\end{align}
where ${g}_{k,i}$ and $\nabla F({w_i})$ are defined as the $i$-th element of $\{{\bf g}_k\}$ and $\nabla F({\bf w}) $, respectively, ${\bm\sigma}=[\sigma_1,\cdots,\sigma_q]$ is a vector of non-negative constants, and the denominator $m_b$ accounts for the fact that the local gradient estimate is computed over a mini-batch of data with size $m_b$.% thus the resultant gradient variance is reduced from $\left\|{\bm\sigma}\right\|_2^2$ to $\frac{\left\|{\bm\sigma}\right\|_2^2}{m_b}$.
 }
\end{assumption}

Notice that the following convergence analysis and power control optimization in Sections \ref{Sec_Con} and \ref{Sec_Power} are based on Assumptions \ref{Assump_Smooth}-\ref{Assum_VarianceBound}, similarly as in prior work \cite{JRen20,DLiu2020Ar,JZhang2021Ar,SXia2020Ar,SWang2021Ar}.
Nevertheless, as shown in simulations in Section~\ref{Sec_CNN}, the proposed power control designs can still work well for CNN when such assumptions are relaxed.

%\vspace{-0.45cm}
\subsection{Over-the-Air Aggregation for FEEL}\label{CommunicationModel}
%\vspace{-0.2cm}

The distributed training latency for FEEL is dominated by the update aggregation process, especially when the number of devices becomes large. Therefore, we focus on the aggregation process over a MAC.
To accelerate the learning, we employ the AirComp technique for fast gradient aggregation by exploiting the superposition property of the MAC. To implement AirComp, during the gradient-uploading phase, all devices transmit simultaneously over the same time-frequency block with proper phase compensation. For ease of exposition, it is assumed that the channel coefficients remain unchanged within each communication round, but may change over different rounds. It is also assumed that the edge devices can perfectly know their own {\it channel state information} (CSI), so they can compensate for the channel phase differences.

Let $\hat h_k^{(n)}$ denote the complex channel coefficient from edge device $k$ to the edge server at communication round $n$, and $h_k^{(n)}$ denote its post-compensated real-valued channel coefficient, i.e., $h_k^{(n)}=|\hat h_k^{(n)}|$.
%For simplicity, we assume symbol-level synchronization among the devices which transmit simultaneously at each communication round.
Then, the received aggregated signal via AirComp (after phase compensation) is given by
 \begin{align}\label{sys_ReceivedSignal}
 	{\bf y}^{(n)}=\sum\limits_{k\in\mathcal K}h_k^{(n)}\sqrt{p_k^{(n)}}{\bf g}_{k}^{(n)}+{\bf z}^{(n)},
 \end{align}
 in which  $p_k^{(n)}$ denotes the power scaling factor at edge device $k$, and ${\bf z}^{(n)}\in\mathbb{R}^q$ denotes the {\it additive white Gaussian noise} (AWGN) with ${\bf z}^{(n)}\sim\mathcal{CN}(0,\sigma_z^2\bf I)$ and  $\sigma_z^2$ being the noise power.
 Based on \eqref{sys_ReceivedSignal}, the global gradient estimate at the edge server is given by\footnote{Unlike the conventional AirComp, using an additional scaling factor at the receiver, in \eqref{sys_ComGlobalGradient} we directly use ${\bf y}^{(n)}/K$ as the estimated value of global gradient for  Air-FEEL. This is due to the fact that the learning rate $\eta^{(n)} $ in \eqref{sys_ModelUpdate} can play the equivalent role of scaling factor, and thus dedicated scaling factors are not needed as in conventional AirComp. }
\begin{align}\label{sys_ComGlobalGradient}
	\hat{\bf g}^{(n)}=\frac{{\bf y}^{(n)}}{K}.
\end{align}
%{\color{blue}
It thus follows from \eqref{sys_ReceivedSignal} and \eqref{sys_ComGlobalGradient} that the aggregation error caused by the over-the-air aggregation in global gradient estimation is given by
\begin{align}\label{sys_Err}
{\bm \varepsilon}^{(n)}&=\hat{\bf g}^{(n)}-\bar{\bf g}^{(n)}
\notag\\
&=\underbrace{\frac{ 1}{K}\sum\limits_{k\in\mathcal K}\left(h_k^{(n)}\sqrt{p_k^{(n)}}-1\right){\bf g}_{k}^{(n)} }_{\rm Signal~misalignment ~error}+\underbrace{\frac{ {\bf z}^{(n)}}{K}}_{\rm Noise},
\end{align}
 which consists of two components representing the signal misalignment error and noise-induced error, respectively.

The devices can adaptively adjust their transmit powers by controlling $\{p_k^{(n)}\}$ to reduce the aggregation errors for enhancing the learning performance.
We consider that each edge device $k\in\mathcal K$ is subject to a maximum power budget $\hat{P}^{\rm max}_k$ for each communication round, and an average power budget denoted by $\hat{P}^{\rm ave}_k$ over the whole training period. Therefore, we have 
\begin{align}
\frac{1}{q}\mathbb{E}\left(\left\|\sqrt{p_{k}^{(n)}}{\bf g}_{k}^{(n)}\right\|^2\right) \leq \hat{P}^{\rm max}_k,~\forall k\in{\mathcal K}, ~\forall n\in\mathcal{N},\label{sys_bar_P_max1}
\end{align}
where $q$ is the size of the gradient vector $g_k^{(n)}$, as well as  
%and $\hat{P}^{\rm max}_k$ denotes the maximum power budget for each single symbol.
%In addition, each device $k\in\mathcal K$ is also constrained by an average power budget denoted by $\hat{P}^{\rm ave}_k$ over the whole training period as expressed below:
\begin{align}
\frac{1}{Nq}\sum \limits_{n\in\mathcal{N}} \mathbb{E}\left(\left\|\sqrt{p_{k}^{(n)}}{\bf g}_{k}^{(n)}\right\|^2\right) \leq \hat{P}^{\rm ave}_k,~\forall k\in{\mathcal K},\label{sys_bar_P_ave1}
\end{align}
where $\mathcal{N}\triangleq\{1,\cdots,N\}$ with $N$ denoting the total number of communication rounds for model training.
%Here, we generally have $\tilde{P}_{k} \le \bar{P}_{k},~\forall k\in{\mathcal K}$.

In the following Section \ref{Sec_Con}, we will establish a direct learning performance metric, namely the optimality gap, linking with the aggregation errors over communication rounds.
Based on the analysis,  in Section \ref{Sec_Power} we will propose to minimize the optimality gap via optimizing the power control subject to a set of individual maximum and average power constraints.

 \section{Convergence Analysis}\label{Sec_Con}

 In this section, we present a convergence analysis framework for the FEEL in the presence of aggregation errors by using the optimality gap as the performance metric, which sheds light on how the imperfect gradient updates affect the convergence of FEEL in general. As will be shown shortly, depending on whether the aggregated gradient estimate is unbiased or not, the FEEL will have different convergence behaviors.

 \subsection{Optimality Gap versus Aggregation Errors}\label{Sec_Error}
% \subsection{}

%In this subsection, the effect of aggregation error caused by the over-the-air aggregation is characterized. 

Suppose that at each communication round $n$, $F\left({\bf w}^{(n)}\right)$ is the value of loss function w.r.t. the parameter vector ${\bf w}^{(n)}$.
% For notational convenience, we use $F\left({\bf w}^{(N+1)}\right)$ to represent $F\left({\bf w}^{(n+1)}\right)$.
%Recall that  ${\bm \varepsilon}^{(n)}$ represent the induced random aggregation error defined in \eqref{sys_Err} at each communication round $n$.
Thus, with the lossy gradient aggregation in \eqref{sys_Err}, the update of learning model at communication round $n$ in \eqref{sys_ModelUpdate} is represented as
\begin{align}\label{Error_ModelUpdate}
{\bf w}^{(n+1)}={\bf w}^{(n)}-\eta^{(n)}\cdot \left(\bar{\bf g}^{(n)}+{\bm \varepsilon}^{(n)}\right),
\end{align}
where ${\bm \varepsilon}^{(n)}$ represents the induced random aggregation error (including the signal misalignment error and noice-induced error) at each communication round $n$.
Let $\mathbb{E}[{\bm \varepsilon}^{(n)}]$ and $\mathbb{E}[\|{\bm \varepsilon}^{(n)}\|^2]$ denote the bias and MSE of the global gradient estimate at each communication round $n$, respectively, where the expectation operation is taken over the stochastic sample selection on the local gradient estimation over a mini-batch dataset, as well as the receiver noise due to AirComp.

Depending on the value of $\mathbb{E}[{\bm \varepsilon}^{(n)}]$, we define two cases for the gradient aggregation.
\begin{itemize}
	\item Case I without unbiased aggregation constraints: The aggregation can either be biased (i.e., $\mathbb{E}[{\bm \varepsilon}^{(n)}]\neq 0$) or unbiased ($\mathbb{E}[{\bm \varepsilon}^{(n)}]= 0$). In this case, no additional constraints on the aggregation biasness are introduced during the power control designs. 
	\item  Case II with unbiased aggregation constraints: The aggregation is unbiased, i.e., the constraints $\mathbb{E}[{\bm \varepsilon}^{(n)}]= 0, \forall n\in\mathcal{N}$, are introduced in the aggregation designs (e.g., transmission power control). 
\end{itemize}

Define the optimality gap after $N$ communication rounds as $F\left({\bf w}^{(N+1)}\right)-F^{\star}$ and  $L\triangleq \|\bf L\|_{\infty}$.
Then, by considering a properly chosen fixed learning rate, we establish the following theorem.

\begin{theorem}[Impact of aggregation error on convergence with a fixed learning rate]\label{Theo_OG_FixedRate}
\emph{ Under Assumption \ref{Assump_Smooth}, suppose that the FEEL algorithm is implemented with a fixed learning rate $\eta\triangleq \eta^{(n)}, \forall n\in\cal N$, with $ 0\leq\eta\leq\frac{2}{2+L}\leq \frac{1}{\delta}$ and fixed mini-batch size $m_b=N$ \cite{signSGD}.
Then, the expected optimality gap satisfies the inequality \eqref{Gap_Iterate_Fix}, where $C=1-\delta\eta$ with $0<C<1$.
\begin{figure*}
	\begin{align}\label{Gap_Iterate_Fix}
&\mathbb{E}\left[F\left({\bf w}^{(N+1)}\right)\right]-F^{\star}\leq \!\underbrace{\sum \limits_{n\in\mathcal{N}}\frac{C^{N-n}}{2} \| \underbrace{\mathbb{E}\left[{\bm \varepsilon}^{(n)}\right]}_{\text{Bias}}\|^2}_{\text{Error floor}} + \notag\\
&~~~~~~~~~~~~~~~~\underbrace{C^N\!\!\underbrace{\left(\mathbb{E}\left[F\left({\bf w}^{(1)}\right)\right]-F^{\star}\right)}_{\text{Initial optimality gap}}\!+\! \!\sum \limits_{n\in\mathcal{N}}\!\frac{C^{N-n}}{2}\left(\underbrace{\frac{ \eta^2 L\left\|{\bm\sigma}\right\|_2^2}{2\delta N K^2}}_{\text{Gradient variance}}\!\!+
\eta^2L^2\leq \| \underbrace{\mathbb{E}\left[{\bm \varepsilon}^{(n)}\right]}_{\text{Bias}}\|^2\!+\!\eta^2L \underbrace{\mathbb{E}\left[\left\|{\bm \varepsilon}^{(n)}\right\|^2\right]}_{\text{MSE}}\!\right)}_{\text{The gap to the error floor}~ \Delta(N)}.
\end{align}
\end{figure*}
}
\end{theorem}
\begin{IEEEproof}
See Appendix~\ref{Theo_OG_FixedRate_Proof}.
\end{IEEEproof}
%{\color{blue}

\begin{figure}
\centering
    \includegraphics[width=3.5in]{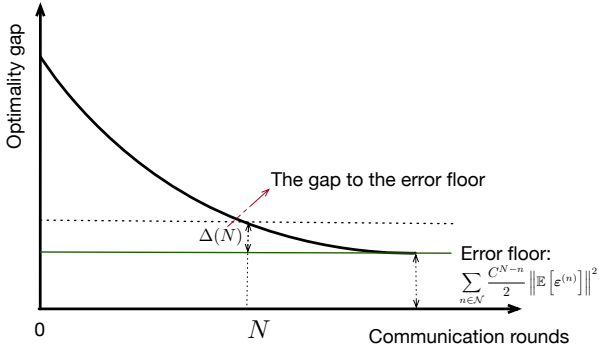}
\caption{Illustration of contraction region on the learning performance. } \label{fig:Contraction}
%-\vspace{-0.4cm}
\end{figure}

\begin{remark}\label{Remark_ContractionRegion}%[Contraction inequality]
\emph{ From Theorem \ref{Theo_OG_FixedRate}, we have the following observations. 
%\footnote{For brevity, the discussion below targets the Theorem \ref{Theo_OG_FixedRate} for the case with fixed learning rate, while we remark that the same insights also hold for the case with diminishing learning rate presented in Corollary \ref{Theo_OG_DRate}.}
\begin{itemize}
\item {\bf The FEEL algorithm converges eventually as $N \rightarrow \infty$, with the optimality gap possibly landed on an error floor instead of diminishing to zero. }
It is observed from \eqref{Gap_Iterate_Fix} that the upper bound of the optimality gap can be decomposed into two components, i.e., the error floor $\sum \limits_{n\in\mathcal{N}}\frac{C^{N-n}}{2}\left\|\mathbb{E}\left[{\bm \varepsilon}^{(n)}\right]\right\|^2$ that cannot vanish as $N$ grows, and the gap to the error floor, denoted by $\Delta(N)$, which can approach zero as $N$ increases. To see this, $\Delta(N)$ is observed to contain four terms related to the initial optimality gap ($\mathbb{E}\left[F\left({\bf w}^{(1)}\right)\right]-F^{\star}$), the gradient variance $\frac{ \eta^2 L\left\|{\bm\sigma}\right\|_2^2}{2\delta N K^2} $, as well as the bias $\mathbb{E}\left[{\bm \varepsilon}^{(n)}\right]$ and MSE $\mathbb{E}\left[\left\|{\bm \varepsilon}^{(n)}\right\|^2\right]$ of the aggregation errors, respectively. All the four terms diminish as $N$ goes to infinity, or become negligible under a sufficiently small learning rate. 
On the other hand, the error floor is determined by the accumulated bias $\left\|\mathbb{E}\left[{\bm \varepsilon}^{(n)}\right]\right\|^2$ over rounds.  Hence, as $N$ increases, the error floor would approach a constant while the gap to it $\Delta(N)$ would vanish. 
 Such effects are illustrated in Fig.~\ref{fig:Contraction}.
\item {\bf The FEEL algorithm shows different convergence behaviors depending on whether the gradient aggregation is biased or not. } For the case with unbiased aggregation constraints, i.e., $\mathbb{E}\left[{\bm \varepsilon}^{(n)}\right]=0, \forall n\in\mathcal{N}$, as $N$ becomes sufficiently large,  the model under training can converge exactly to the optimal point with minimum training loss with zero error floor in the training process. By contrast, for the case without unbiased aggregation constraints,  the model under training may only converge to a neighborhood of the optimal point (if the aggregation is biased). However, the case with unbiased aggregation constraints may converge slower compared with its counterpart without unbiased aggregation constraints, as the enforcement of the unbiasness generally comes at a cost of elevated MSE that translates to a larger gap to the error floor $\Delta(N)$. The observation is also validated via experiments shown in Section \ref{sec_simu}. 
%\item {\bf There is a tradeoff in designing the learning rate $\eta$ between the error floor and the gap to the error floor. } It is observed that when  the learning rate is small, $\Delta$ trends to be small, whereas the error floor would be large as $\upsilon^{(n)}$ increases. 
\item {\bf Latter rounds are more sensitive to aggregation error.} The bias $\mathbb{E}\left[{\bm \varepsilon}^{(n)}\right]$ and MSE $\mathbb{E}\left[\left\|{\bm \varepsilon}^{(n)}\right\|^2\right]$ at the later communication rounds (with large $n$) contribute more on the optimality gap than that of the initial rounds (with small $n$), as the effect of the aggregation error introduced at early stages (small $n$) is discounted by $C^{N-n}$ on the right hand side of \eqref{Gap_Iterate_Fix}.
% in both consisting components  of the optimality gap, i.e., the error floor and the gap to the error floor. 
\end{itemize}} 
\end{remark}

Theorem \ref{Theo_OG_FixedRate} can be extended to the case with diminishing learning rates, as shown in the following corollary.

\begin{corollary}[Impact of aggregation error on convergence with diminishing learning rates]\label{Theo_OG_DRate}
\emph{ Under Assumption \ref{Assump_Smooth}, suppose that the FEEL algorithm is implemented with fixed mini-batch size $m_b=N$, and diminishing learning rates $\eta^{(n)}=\frac{u}{n+v}, \forall n\in\mathcal{N}$ \cite{Bottou2018}, with $v>0$ and $u>1/\delta$,  such that  $\eta^{(1)} \leq\frac{2}{2+L}$.
Then, the expected optimality gap satisfies the following inequality:
\begin{align}\label{Gap_Iterate}
&\mathbb{E}\left[F\left({\bf w}^{(N+1)}\right)\right]\!-\!F^{\star}\leq \left(\!\prod\limits_{n\in\mathcal{N} }C^{(n)}\!\right)\!\left(\mathbb{E}\left[F\left({\bf w}^{(1)}\right)\right]\!-\!F^{\star}\right)\notag\\
&+\sum_{n=1}^{N}J^{(n)}\left(\frac{(\eta^{(n)})^2L\left\|{\bm\sigma}\right\|_2^2}{2m_bK^2}+(\eta^{(n)})^2L^2\left\|\mathbb{E}\left[{\bm \varepsilon}^{(n)}\right]\right\|^2\right)\notag\\
&\!+\!\sum_{n=1}^{N}J^{(n)}\!\left\|\mathbb{E}\left[{\bm \varepsilon}^{(n)}\right]\right\|^2\!\!+\!\sum_{n=1}^{N}\!J^{(n)}(\eta^{(n)})^2L \mathbb{E}\!\left[\left\|{\bm \varepsilon}^{(n)}\right\|^2\right],
\end{align}
where $C^{(n)}=1-\delta\eta^{(n)}$ and  $ J^{(n)}\triangleq\frac{\prod_{i=n}^{N}C^{(i)}}{2C^{(n)}}, \forall n\in\mathcal{N} $.%$\theta=\frac{L\left\|{\bm\sigma}\right\|_2^2}{2m_bK^2}$, 
}\!
\end{corollary}
\begin{IEEEproof}
This proof is similar to that for Theorem \ref{Theo_OG_FixedRate}, and thus is omitted here for brevity.
\end{IEEEproof}

Note that similar observations can be made from Corollary \ref{Theo_OG_DRate} as those in Remark 1.  An additional tradeoff lies in designing the learning rate $\eta^{(n)}$.  It is observed that when the learning rate is small, the error floor would be large as $C^{(n)}$ (or $J^{(n)}$) tends to be increasing, whereas the gap to the error floor becomes small.

%, which are thus omitted for brevity. 

%Note that the same observations in Remark 1 can be made from the following Corollary \ref{Theo_OG_DRate} for the case with diminishing learning rate, which are thus omitted for brevity.

\subsection{Optimality Gap versus Transmission Power Control}
In this subsection, we  obtain the optimality gap w.r.t. the transmission power control variables based on the results in Section~\ref{Sec_Error}, in order to facilitate the power control design in the sequel. In particular, we consider the Air-FEEL in the cases without and with unbiased aggregation constraints $\mathbb{E}\left[{\bm \varepsilon}^{(n)}\right]=0, \forall n\in\mathcal{N}$.

%, namely $\mathbb{E}\left[{\bm \varepsilon}^{(n)}\right]\neq0$ and $\mathbb{E}\left[{\bm \varepsilon}^{(n)}\right]=0, \forall n\in\mathcal{N}$, respectively. 
%On the basic of Theorem~\ref{Theo_OG_DRate}, we can directly extend the convergence analysis with generic aggregation error into the special one owing to Air-FEEL system.  

Before proceeding, we introduce the following assumption on the sample-wise gradient bound.
 \begin{assumption}[Bounded sample-wise gradient]\label{Assump_SamplewiseBound}\emph{
 At any communication round $n$, the sample-wise gradient $\nabla f\left({\bf w}^{(n)},{\bf x},y\right) $ for any training sample $( {\bf x},y)$ is upper bounded by a given constant $G^{(n)}$, i.e.,
 \begin{align}
\left\| \nabla f\left({\bf w}^{(n)},{\bf x},y\right) \right\| \leq G^{(n)}, ~\forall n\in\mathcal{N}.
%\left\|{\bf g}_{k}^{(n)}\right\| \leq G_k, ~\forall k\in\mathcal{K},~n\in\mathcal{N}.
 \end{align}}
\end{assumption}

Based on Assumption \ref{Assump_SamplewiseBound}, we have $\left\|\nabla F({\bf w}^{(n)}) \right\|\leq \max_{ ( {\bf x},y)\in\mathcal{D}}\left\| \nabla f\left({\bf w}^{(n)},{\bf x},y\right) \right\| \leq G^{(n)}$. 
Together with Assumption~\ref{Assum_VarianceBound}, it thus holds that 
\begin{align}\label{Gradient_square}
	 \mathbb{E}\left[\left\|{\bf g}_{k}^{(n)}\right\|^2\right] &\leq \left\|\nabla F({\bf w}^{(n)}) \right\|^2+\frac{\left\|{\bm\sigma}\right\|_2^2}{m_b} \notag\\
	 &\leq \hat{G}^{(n)}\triangleq \left(G^{(n)}\right)^2+\frac{\left\|{\bm\sigma}\right\|_2^2}{m_b} .
\end{align}
%where $\hat{G}^{(n)}$.

\subsubsection{Convergence Analysis for Air-FEEL in Case I}
In this part, we formally characterize the convergence behavior of Air-FEEL  w.r.t. the transmission power in the case without unbiased aggregation constraints.

% with $\mathbb{E}\left[{\bm \varepsilon}^{(n)}\right]\neq0, \forall n\in\cal N$ in general.

According to the definition of $ {\bm \varepsilon}^{(n)}$ in formula \eqref{sys_Err}, at each communication round $n$, the bias and MSE of gradient estimates through the over-the-air gradient aggregation for Air-FEEL are derived as 
\begin{align}
	\!\!\!\!\!\left\|\mathbb{E}\left[{\bm \varepsilon}^{(n)}\right]\right\|&=\frac{ \left\|\nabla F({\bf w}^{(n)}) \right\|}{K}\left(\sum\limits_{k\in\mathcal K}h_k^{(n)}\sqrt{p_k^{(n)}}-K\right)\notag\\
	&\leq \frac{G^{(n)} }{K}\left(\sum\limits_{k\in\mathcal K}h_k^{(n)}\sqrt{p_k^{(n)}}-K\right),\label{equ_bias_p}
	\end{align}
	\begin{align}
	&\!\!\!\!\mathbb{E}\!\left[\!\left\|{\bm \varepsilon}^{(n)}\right\|^2\right]\leq\!\frac{ \!\left\|\nabla F({\bf w}^{(n)}) \right\|^2\!\!+\!\!\frac{\left\|{\bm\sigma}\right\|_2^2}{m_b}}{K}\sum\limits_{k\in\mathcal K}\!\left(\!h_k^{(n)}\sqrt{p_k^{(n)}}\!-\!1\!\right)^2\!\!+\!\frac{\sigma_z^2q}{K^2}\notag\\
	&~~~~~~~~~~~~\leq \frac{ \hat{G}^{(n)}}{K}\sum\limits_{k\in\mathcal K}\left(h_k^{(n)}\sqrt{p_k^{(n)}}-1\right)^2+\frac{\sigma_z^2q}{K^2},\label{equ_unbias_p}
\end{align} 
where both inequalities follow from Assumptions~\ref{Assum_VarianceBound} and \ref{Assump_SamplewiseBound}.
By substituting \eqref{equ_bias_p} and \eqref{equ_unbias_p} into \eqref{Gap_Iterate} and \eqref{Gap_Iterate_Fix}, we can have the following proposition.

%By replacing the squared bias (i.e., $\left\|\mathbb{E}\left[{\bm \varepsilon}^{(n)}\right]\right\|^2$) and MSE (i.e., $ \mathbb{E}\left[\left\|{\bm \varepsilon}^{(n)}\right\|^2\right]$) of the over-the-air gradient aggregation for Air-FEEL system, we can obtain the corresponding convergence behavior under the biased aggregation policy as follows, which is derived as a function of power control.

\begin{proposition}[Optimality gap for Air-FEEL without unbiased aggregation constraints]\label{Theo_OG_DRate_Air}
\emph{ The expected optimality gap for Air-FEEL in the case without unbiased aggregation constraints is upper bounded by
\begin{align}\label{Gap_Iterate_Air}
&\mathbb{E}\left[F\left({\bf w}^{(N+1)}\right)\right]\!-\!F^{\star}\leq \prod\limits_{n\in\mathcal{N} }C^{(n)}\left(\left[F\left({\bf w}^{(1)}\right)\right]-F^{\star}\right)\notag\\
&\!+\!\sum_{n=1}^{N}J^{(n)}\!\left(\!
A^{(n)}\!\left(\!\sum\limits_{k\in\mathcal K}h_k^{(n)}\sqrt{p_k^{(n)}}-K\right)^2\!+\!\frac{(\eta^{(n)})^2L\left\|{\bm\sigma}\right\|_2^2}{2m_bK^2}\!\right)\notag\\
&\!+\!\sum_{n=1}^{N}\!J^{(n)}\!\left(\!B^{(n)}\!\sum\limits_{k\in\mathcal K}\!\left(\!h_k^{(n)}\sqrt{p_k^{(n)}}\!-\!1\!\right)^2\!\!\!+\!\frac{(\eta^{(n)})^2 \sigma_z^2Lq}{K^2}\!\right)\!,\!\!
\end{align}
where $A^{(n)}= \frac{\left(1+(\eta^{(n)})^2L^2\right)(G^{(n)})^2 }{K^2}$, $B^{(n)}=\frac{ (\eta^{(n)})^2L \hat{G}^{(n)}}{K}$, and $ J^{(n)}=\frac{\prod_{i=n}^{N}C^{(i)}}{2C^{(n)}}$ for the diminishing learning rates  $\eta^{(n)}=\frac{u}{n+v}, \forall n\in\mathcal{N}$, with $v>0,u>1/\delta$, and $\eta^{(1)} \leq\frac{2}{2+L}$; while $A^{(n)}= \frac{\left(1+\eta^2L^2\right)(G^{(n)})^2 }{K^2}$, $B^{(n)}=\frac{ \eta^2L \hat{G}^{(n)}}{K}$, and $J^{(n)}=\frac{C^{N-n}}{2}$ for the fixed learning rate with $\eta=\eta^{(n)}, \forall n\in\cal N$, with $ 0\leq\eta\leq\frac{2}{2+L}\leq \frac{1}{\delta}$.
}
\end{proposition}
%\begin{IEEEproof}
%Please refer to Appendix~\ref{Theo_OG_FixedRate_Proof}.
%\end{IEEEproof}

%\begin{proposition}[Optimality gap for Air-FEEL under biased aggregation with a fixed learning rate]\label{Theo_OG_FixedRate_Air}
%\emph{ With a fixed learning rate $\eta$ satisfying $ 0\leq\eta\leq\frac{2}{2+L}\leq \frac{1}{\delta}$, the expected optimality gap for Air-FEEL under biased aggregation is upper bounded by
%\begin{align}
%&\mathbb{E}\left[F\left({\bf w}^{(N+1)}\right)\right]-F^{\star}\notag\\
%&~\leq C^N\left(\mathbb{E}\left[F\left({\bf w}^{(1)}\right)\right]-F^{\star}\right) 
%+ \sum \limits_{n\in\mathcal{N}}\frac{C^{N-n}}{2} \left( 
%\frac{\left(1+\eta^2L^2\right)(G^{(n)})^2 }{K^2}\left(\sum\limits_{k\in\mathcal K}h_k^{(n)}\sqrt{p_k^{(n)}}-K\right)^2+\eta\theta\right)\notag\\
%&~~+\sum \limits_{n\in\mathcal{N}}\frac{C^{N-n}}{2} \left( \frac{\eta^2L \hat{G}^{(n)}}{K^2}\sum\limits_{k\in\mathcal K}\left(h_k^{(n)}\sqrt{p_k^{(n)}}-1\right)^2+\frac{\eta^2\sigma_z^2Lq}{K^2} \right).
%\end{align}}
%\end{proposition}

\subsubsection{Convergence Analysis for Air-FEEL in Case II}
Next, we consider the case with unbiased aggregation constraints, where we have  $\mathbb{E}\left[{\bm \varepsilon}^{(n)}\right]=0, \forall n\in\cal N$. 
Similar to Proposition \ref{Theo_OG_DRate_Air}, we have the following proposition.
%Hence, the corresponding convergence analysis is derived as follows.
\begin{proposition}[Optimality gap for Air-FEEL with unbiased aggregation constraints]\label{Theo_OG_DRate_Air_ub}
\emph{ The expected optimality gap for Air-FEEL in the case with unbiased aggregation constraints is upper bounded by
\begin{align}\label{Gap_Iterate_Air_ub}
&\mathbb{E}\left[F\left({\bf w}^{(N+1)}\right)\right]-F^{\star}\leq \prod\limits_{n\in\mathcal{N} }C^{(n)}\left(\left[F\left({\bf w}^{(1)}\right)\right]-F^{\star}\right) \notag\\
&+\sum_{n=1}^{N}J^{(n)}\left(B^{(n)}\sum\limits_{k\in\mathcal K}\left(h_k^{(n)}\sqrt{p_k^{(n)}}-1\right)^2+\frac{(\eta^{(n)})^2 \sigma_z^2Lq}{K^2}\right)\notag\\
&+\sum_{n=1}^{N}J^{(n)}\frac{(\eta^{(n)})^2L\left\|{\bm\sigma}\right\|_2^2}{2m_bK^2},
\end{align}
where $B^{(n)}=\frac{ (\eta^{(n)})^2L \hat{G}^{(n)}}{K^2}$ and $ J^{(n)}=\frac{\prod_{i=n}^{N}C^{(i)}}{2C^{(n)}}$ for the diminishing learning rates  $\eta^{(n)}=\frac{u}{n+v}, \forall n\in\mathcal{N}$, with $v>0,u>1/\delta$, and $\eta^{(1)} \leq\frac{2}{2+L}$; while $B^{(n)}=\frac{ \eta^2L \hat{G}^{(n)}}{K^2}$ and $J^{(n)}=\frac{C^{N-n}}{2}$ for the fixed learning rate with $\eta=\eta^{(n)}, \forall n\in\cal N$, with $ 0\leq\eta\leq\frac{2}{2+L}\leq \frac{1}{\delta}$.
}
\end{proposition}

%%Note that Propositions \ref{Theo_OG_DRate_Air} and \ref{Theo_OG_DRate_Air_ub} could be reduced to that for the case with fixed learning rate via $\eta^{(n)}=\eta, \forall n\in\cal N$, with $ 0\leq\eta\leq\frac{2}{2+L}\leq \frac{1}{\delta}$. 
%Notice that the derived convergence results under the fixed learning rate has a similar form with that under diminishing learning rates.
%Thus, the subsequent power control optimization will be presented targeting the case with diminishing learning rates only for brevity, while the yielded insights hold for both cases. 

Since the derived convergence results for both the cases of diminishing and fixed learning rates share similar form, the subsequent power control optimization will be presented targeting the case with diminishing learning rates only for brevity, while the yielded insights hold for both cases.

 \section{Power Control Optimization}\label{Sec_Power}

 Given the convergence results of Air-FEEL in the preceding section, we are now ready to present the power control optimization polices  for speeding up the convergence rate in this section. 
%Particularly, we formulate the problem to minimize the optimality gap derived in Propositions \ref{Theo_OG_DRate_Air} and \ref{Theo_OG_DRate_Air_ub} via power control under a set of long-term and maximum power constraints imposed on individual devices. The problems will be presented and solved for both cases of biased and unbiased aggregation respectively in the sequel.  

To start with, we first reformulate the power constraints in \eqref{sys_bar_P_max1} and \eqref{sys_bar_P_ave1} by leveraging Assumption~\ref{Assump_SamplewiseBound} and inequality \eqref{Gradient_square} to avoid the requirement of non-causal gradient information $g_k^{(n)}$.
Hence, the individual power constraints at each communication round and the entire training process are respectively reformulated as
\begin{align}
&p_{k}^{(n)} \hat{G}^{(n)}\leq P^{\rm max}_k,~\forall k\in{\mathcal K}, ~ n\in\mathcal{N},\label{sys_bar_P_max}\\
&\frac{1}{N}\sum \limits_{n\in\mathcal{N}}p_{k}^{(n)}\hat{G}^{(n)}\leq P^{\rm ave}_k,~\forall k\in{\mathcal K},\label{sys_bar_P_ave}
\end{align} 
where $P^{\rm max}_k\triangleq q\hat{P}^{\rm max}_k$ and $P^{\rm ave}_k\triangleq q\hat{P}^{\rm ave}_k, \forall k\in\mathcal{K},$ are defined for notational convenience.

 \subsection{Power Control Optimization  for  Case I}
 
We start with the case I without unbiased aggregation constraints.
Discarding the irrelevant terms in \eqref{Gap_Iterate_Air} in Proposition \ref{Theo_OG_DRate_Air} (i.e., the terms related to the initial optimality gap $\mathbb{E}\left[F\left({\bf w}^{(1)}\right)\right]-\!F^{\star}$, the gradient variance bound $\frac{(\eta^{(n)})^2L\left\|{\bm\sigma}\right\|_2^2}{2m_bK^2}$, and the noise power $\frac{(\eta^{(n)})^2 \sigma_z^2Lq}{K^2}$) in Proposition ~\ref{Theo_OG_DRate_Air}, we denote $\tilde{\Phi}(\{p_k^{(n)}\})$ in the following as the effective optimality gap to be optimized. 
\begin{align}\label{Eff_OG_Biased}
\tilde{\Phi}(\{p_k^{(n)}\}) \triangleq &\sum_{n=1}^{N}J^{(n)}A^{(n)}\left(\sum\limits_{k\in\mathcal K}h_k^{(n)}\sqrt{p_k^{(n)}}-K\right)^2\notag\\
&+\sum_{n=1}^{N}J^{(n)}B^{(n)}\sum\limits_{k\in\mathcal K}\left(h_k^{(n)}\sqrt{p_k^{(n)}}-1\right)^2.
\end{align}
The optimization problem is thus formulated as
  \begin{align}
 	\mathbf{P1:} \min_{\{p_k^{(n)}\ge 0\}} ~~& \tilde{\Phi}(\{p_k^{(n)}\})\notag\\
 {\rm s.t.}~~~~&\eqref{sys_bar_P_max}~\text{and}~\eqref{sys_bar_P_ave}.\notag
 \end{align}
%Note that the power scaling factors at different devices are coupled with each other in the objective function in \eqref{Eff_OG_Biased}, leading to a non-convex problem. 
By introducing  a set of auxiliary variables, $\hat{p}_k^{(n)}=\sqrt{p_k^{(n)}}, \forall k\in\mathcal{K}, n\in\mathcal{N}$, the objective is re-expressed as 
\begin{align}\label{Eff_OG_Biased_Phat}
{\Phi}(\{\hat{p}_k^{(n)}\})\triangleq \sum_{n=1}^{N}J^{(n)}A^{(n)}\left(\sum\limits_{k\in\mathcal K}h_k^{(n)}\hat{p}_k^{(n)}-K\right)^2\notag\\
+\sum_{n=1}^{N}J^{(n)}B^{(n)}\sum\limits_{k\in\mathcal K}\left(h_k^{(n)}\hat{p}_k^{(n)}-1\right)^2\!\!,\!
\end{align}
and problem (P1) is re-expressed as 
\begin{align}
 \!\!\!\mathbf{P1.1:} \min_{\{\hat{p}_k^{(n)}\ge 0\}} ~&{\Phi}(\{\hat{p}_k^{(n)}\})\notag\\
 {\rm s.t.}~~~~ &\hat{q}_{k}^{(n)} \leq P^{\rm max}_{k,n},~\forall k\in{\mathcal K}, ~ n\in\mathcal{N} \label{Biased_MaxPower}\\
&\frac{1}{N}\sum \limits_{n\in\mathcal{N}}\left(\hat{q}_{k}^{(n)}\right)^2\hat{G}^{(n)} \leq P^{\rm ave}_k,~\forall k\in{\mathcal K}\label{Biased_AvePower},
\end{align}
where constraints \eqref{Biased_MaxPower} and \eqref{Biased_AvePower} follow from \eqref{sys_bar_P_max} and \eqref{sys_bar_P_ave}, respectively, and $P^{\rm max}_{k,n}\triangleq \sqrt{\frac{ P^{\rm max}_k}{\hat{G}^{(n)}}}, \forall k\in\mathcal{K},~n\in\mathcal{N}$. Problem (P1.1) is convex and can thus be optimally solved by the standard convex optimization techniques such as the interior point method \cite{cvx}. Alternatively, to gain engineering insights, we resort to the Lagrange duality method to derive the structured optimal solution for problem (P1.1). Let $\{\hat{p}_k^{(n)\rm opt}\}$ denote the optimal solution to problem (P1.1), and $\varphi_k^{\rm opt}, \forall k\in\mathcal{K}$ the optimal dual variable  associated with the $k$-th constraint in \eqref{Biased_AvePower}. Then we have the following proposition. 

\begin{proposition}\label{lemma_Biased_Power}\emph{
The optimal solution  $\hat{p}_k^{(n)\rm opt}, \forall k\in\mathcal{K},~n\in\mathcal{N}$ to problem (P1.1) is 
\begin{align}\label{Biased_hatP_Opt}
\hat{p}_k^{(n)\rm opt}=\min\left[\frac{B^{(n)}+A^{(n)}K }{M_k^{(n)}+ A^{(n)}M_k^{(n)}\sum\limits_{i\in\mathcal{K}} \frac{h_i^{(n)}}{M_i^{(n)}} } ,P^{\rm max}_{k,n}\right],
\end{align}
where $ M_k^{(n)}\triangleq B^{(n)} h_k^{(n)}+\frac{ \varphi_k^{\rm opt}\hat{G}^{(n)}}{NJ^{(n)}h_k^{(n)}} , \forall k\in\mathcal{K},~n\in\mathcal{N} $.
	}
\end{proposition}
\begin{IEEEproof}
See Appendix~\ref{Proof_Lemma_Biased}.
\end{IEEEproof}

\noindent According to Proposition~\ref{lemma_Biased_Power}, the optimal power scaling factors $p_k^{(n)\rm opt}, \forall k\in\mathcal{K},~n\in\mathcal{N}$ to problem (P1) is 
\begin{align}\label{Biased_power_Opt}
	\!\!\!p_k^{(n)\rm opt}\!=\!\min\!\!\left[\!\left(\frac{B^{(n)}+A^{(n)}K }{M_k^{(n)}\!+\! A^{(n)}M_k^{(n)}\!\!\sum\limits_{i\in\mathcal{K}}\!\! \frac{h_i^{(n)}}{M_i^{(n)}} } \right)^2\!\!\!\!,\!\left(P^{\rm max}_{k,n}\right)^2\right]\!\!.\!
\end{align}
%From \eqref{Biased_power_Opt}, the optimal power scaling factors among devices should related to all the average constraints and the maximum power constraints at each communication rounds.

\begin{remark}\emph{According to Proposition \ref{lemma_Biased_Power}, the optimal $\{\hat{p}_k^{(n)\rm opt}\}$ (equivalently the optimal power scaling factor $p_k^{(n)\rm opt}=(\hat{p}_k^{(n)\rm opt})^2, \forall k\in\mathcal{K},~n\in\mathcal{N}$) exhibits a {\it regularized channel inversion} structure with the regularized term $\sum\limits_{i\in\mathcal{K}} \frac{A^{(n)}h_i^{(n)}M_k^{(n)}}{M_i^{(n)}}$ related to all dual variables $ \varphi_k^{\rm opt}$ associated with the average power budgets  at all edge devices in \eqref{Biased_AvePower}. Considering the special case when the average power budgets $\{P^{\rm ave}_k\}$ at all devices are sufficiently large, such that the dual variables become zero at the same time (i.e., $\varphi_k^{\rm opt}=0, \forall k\in\mathcal{K},~n\in\mathcal{N}$).  In this case, the optimal power scaling strategy reduces to the channel inversion policy, i.e., $p_k^{(n)\rm opt}=\min\left[\frac{1}{\left(h_k^{(n)}\right)^2}, \left(P^{\rm max}_{k,n}\right)^2\right]$, $\forall k\in\mathcal{K},~n\in\mathcal{N}$.
Interestingly, this result is equivalent to minimizing the MSE in isolation at each communication round.
In other words,  in the special case when all devices have a sufficiently large average power budgets,  the conventional MSE minimization can be sufficient to minimize the optimality gap. 
	}
\end{remark}

 \subsection{Power Control Optimization  fo Case II }
 
Next, we consider the power control optimization for the case with unbiased aggregation constraints, where the power control policy needs to enforce the additional constraint $\mathbb{E}\left[{\bm \varepsilon}^{(n)}\right]= 0, \forall n\in\cal N$.  
 According to \eqref{equ_bias_p}, it follows that  $\sum\limits_{k\in\mathcal K}h_k^{(n)}\sqrt{p_k^{(n)}}=K, \forall n\in\cal N$. 
In this case, the effective optimality gap in Proposition~\ref{Theo_OG_DRate_Air_ub} is given by
\begin{align}
\tilde\Theta\left(\{p_k^{(n)}\}\right)\triangleq \sum_{n=1}^{N}J^{(n)}B^{(n)}\sum\limits_{k\in\mathcal K}\left(h_k^{(n)}\sqrt{p_k^{(n)}}-1\right)^2\!.
\end{align}
Accordingly, we formulate the power control optimization problem as 
\begin{align}
	 	\mathbf{P2:} \min_{\{p_k^{(n)}\ge 0\}} ~~&\tilde\Theta\left(\{p_k^{(n)}\}\right)\notag\\
 {\rm s.t.}~~~~&\sum\limits_{k\in\mathcal K}h_k^{(n)}\sqrt{p_k^{(n)}}=K, \forall n\in\cal N\\
 &\eqref{sys_bar_P_max}~\text{and}~\eqref{sys_bar_P_ave}.\notag
\end{align}
Note that problem (P2) is non-convex. However, via a change of variables $ q_k^{(n)}\triangleq\sqrt{p_k^{(n)}}, \forall k\in\mathcal{K}, n\in\mathcal{N}$, the objective can be re-expressed as 
\begin{align}
\Theta\left(\{q_k^{(n)}\}\right)\triangleq \sum_{n=1}^{N}J^{(n)}B^{(n)}\sum\limits_{k\in\mathcal K}\left(h_k^{(n)}q_k^{(n)}-1\right)^2,
\end{align}
%where $\Theta\left(\{q_k^{(n)}\}\right)$ is convex w.r.t. $ q_k^{(n)}$.
and problem (P2) can be transformed into the following equivalent convex form: 
\begin{align}
 	\!\!\!\mathbf{P2.1:} \min_{\{q_k^{(n)}\ge 0\}} ~&\Theta\left(\{q_k^{(n)}\}\right)\notag\\
 {\rm s.t.}~~~&\sum\limits_{k\in\mathcal K}h_k^{(n)}q_{k}^{(n)}=K,~\forall n\in\mathcal{N} \label{Unbiased_alignment}\\
 &q_{k}^{(n)} \leq P^{\rm max}_{k,n},~\forall k\in{\mathcal K}, ~\forall n\in\mathcal{N} \label{Unbiased_MaxPower}\\
&\frac{1}{N}\sum \limits_{n\in\mathcal{N}}\left(q_{k}^{(n)}\right)^2\hat{G}^{(n)} \leq P^{\rm ave}_k,~\forall k\in{\mathcal K}\label{Unbiased_AvePower},
\end{align}
where constraints \eqref{Unbiased_MaxPower} and \eqref{Unbiased_AvePower} follow from \eqref{sys_bar_P_max} and \eqref{sys_bar_P_ave}, respectively.
%Notice that problem (P2.1) is convex and thus can be optimally solved.

\subsubsection{Feasibility of Problem (P2.1)} 
Before solving problem (P2.1), we first check its feasibility, i.e., whether the power budget can support the required unbiased estimation level denoted by $\ell$ or not. 
Let $\ell^{\star} $ denote the maximum unbiased estimation level, which can be expressed as 
\begin{align}
\ell^{\star}= \max_{\{q_k^{(n)}\ge 0\}} ~~&\ell \label{Unbiased_Fea}\\
 {\rm s.t.}~~~~&\sum\limits_{k\in\mathcal K}h_k^{(n)}q_{k}^{(n)}\geq \ell,~\forall n\in\mathcal{N}\notag\\
 &\eqref{Unbiased_MaxPower}~\text{and }~\eqref{Unbiased_AvePower}.\notag
\end{align}
If $\ell^{\star}\geq K$, then problem (P2.1) is feasible; otherwise,  problem (P2.1) is not feasible. 
Hence, the feasibility checking procedure corresponds to finding $\ell^{\star} $ by solving problem \eqref{Unbiased_Fea}.
Notice that problem \eqref{Unbiased_Fea} is convex, which can thus be efficiently solved via standard convex optimization techniques, such as the interior point method \cite{cvx}. By comparing $\ell^{\star}$ versus $K$, the feasibility of problem (P2.1) is checked. In the following, we solve problem (P2.1) when it is feasible.

\subsubsection{Optimal Solution to Problem (P2.1)} 
 Let $\{q_k^{(n)\rm opt}\}$ denote the optimal solution to problem (P2.1).
We have the following proposition by leveraging the Lagrange duality method, where $\mu_n^{\rm opt} $  and $\lambda_k^{\rm opt}$ are the optimal dual variables associated with constraints \eqref{Unbiased_alignment} and \eqref{Unbiased_AvePower}, respectively.

\begin{proposition}\label{theorem_Unbiased_gamma}\emph{The optimal solution $q_k^{(n)\rm opt}, \forall k\in\mathcal{K},~n\in\mathcal{N} $ to problem (P2.1) is given as
\begin{align}
q_k^{(n)\rm opt}=\min\left[\frac{h_k^{(n)}\alpha_k^{(n)}}{ (h_k^{(n)})^2+\frac{ 2\lambda_k^{\rm opt}\hat{G}^{(n)}}{NJ^{(n)}B^{(n)}}},P^{\rm max}_{k,n}\right],\label{Unbiased_q_Opt}
\end{align}
where $\alpha_k^{(n)}\triangleq  \left(1-\frac{\mu_n^{\rm opt}}{2J^{(n)}B^{(n)}} \right)^+, \forall k\in\mathcal{K},~n\in\mathcal{N}$.
}
\end{proposition}
\begin{IEEEproof}
See Appendix~\ref{Proof_theorem_Unbiased_gamma}.
\end{IEEEproof}

From \eqref{Unbiased_q_Opt} in Proposition~\ref{theorem_Unbiased_gamma}, we can accordingly obtain the optimal power scaling factors $p_k^{(n)\rm opt}, \forall k\in\mathcal{K},~n\in\mathcal{N}$ to problem (P2) as
\begin{align}\label{Unbiased_power_Opt}
	p_k^{(n)\rm opt}&=\left(q_k^{(n)\rm opt}\right)^2\notag\\
	&=\!\min\!\left[\!\left(\!\frac{h_k^{(n)}\alpha_k^{(n)}}{ (h_k^{(n)})^2\!+\!\frac{ 2\lambda_k^{\rm opt}\hat{G}^{(n)}}{NJ^{(n)}B^{(n)}}}\right)^2\!,\left(P^{\rm max}_{k,n}\right)^2\!\right]\!.
\end{align}

\begin{remark}\emph{According to \eqref{Unbiased_q_Opt}, the optimal solution of $\{q_k^{(n)\rm opt}\}$ to problem (P2.1) (equivalently the optimal power scaling factor $p_k^{(n)\rm opt}=(q_k^{(n)\rm opt})^2, \forall k\in\mathcal{K},~n\in\mathcal{N}$) has a similar \emph{regularized channel inversion} structure as that in \eqref{Biased_hatP_Opt}, but the regularized term therein (i.e.,$\frac{ 2\lambda_k^{\rm opt}\hat{G}^{(n)}}{NJ^{(n)}B^{(n)}}$) is {\it only} related to its own device $k$'s average power budget in \eqref{Unbiased_AvePower} through the dual variable $\lambda_k^{\rm opt}$, as opposed to all devices' budgets in \eqref{Biased_AvePower} for the case without unbiased aggregation constraints. 
Furthermore, it is observed that for any edge device $k\in\mathcal{K}$, if $\lambda_k^{\rm opt}>0$ holds, then the average power constraint of edge device $k$ must be tight at the optimality  (i.e., $\frac{1}{N}\sum \limits_{n\in\mathcal{N}}\left(q_{k}^{(n)\rm opt}\right)^2\hat{G}^{(n)} - P^{\rm ave}_k=0$) due to the complementary slackness condition, and thus this edge device should use up its average power budget based on the regularized channel inversion power control over communication rounds; otherwise, if $\lambda_k^{\rm opt}=0$, then edge device $k$ should transmit with channel-inversion power control without using up its average power budget.
%Note that the regularization channel inversion strategy in unbiased aggregation case is quite different from that for the biased aggregation case, where the regularization term at each edge device is additionally related to all other devices' average power budgets.
}
\end{remark}

\section{Simulation Results}\label{sec_simu}
%linear regression and handwritten digit identification
In this section, we provide simulation results to validate the performance of the proposed power control policies for Air-FEEL. 
The proposed algorithms are implemented using the Matlab and Pytorch for two different tasks, i.e., the ridge regression and handwritten digit recognition, respectively.

\subsection{Simulation Setup and Benchmark Schemes}
In the simulation, the wireless channels from the edge devices to the edge server over different communication rounds follow i.i.d. Rayleigh fading, i.e., $h_k^{(n)}$'s are modeled as i.i.d. {\it circularly symmetric complex Gaussian} (CSCG) random variables with zero mean and unit variance. 
%We set the received {\it signal-to-ratio ratio} (SNR) at the edge server as $\frac{P^{\rm ave}\mathbb{E}[|h_k^{(n)}|^2]}{\sigma_z^2q}=10$ dB with $\hat{P}^{\rm ave}=\sum\limits_{k\in\mathcal{K}}\hat{P}^{\rm ave}_k/K$, where
We set the number of devices as $K=10$, the noise variance $\sigma_z^2=0.1$, and the average power budgets at different devices $\hat{P}^{\rm ave}_k$ to be heterogeneous\footnote{The average power budgets at different devices are set as, $\hat{P}^{\rm ave}_i=5$W and  $\hat{P}^{\rm ave}_{i+1}=15$W,  $i=\{1,\cdots,K/2\}$.}.
We set the maximum power budget to be $5\hat{P}^{\rm ave}$. 
We consider both the fixed and diminishing learning rates with $\eta=0.05$ and $\eta^{(n)}=\frac{u}{n+v}$ under $u=2$ and $v=8$, respectively. 
As for the performance metrics,  the optimality gap and prediction error are considered for ridge regression on synthetic dataset, while the loss function value and test (recognition) accuracy are considered for handwritten digit recognition on MNIST dataset.

For performance comparison, we consider the following two benchmark schemes.
\begin{itemize}
	\item  {\bf Fixed power transmission}: The edge devices transmit with fixed power over different communication rounds by setting $p_k^{(n)}=P^{\rm ave}_k, \forall k\in\cal K$.
	\item {\bf Conventional MSE minimization}: The edge devices optimize their  power control to minimize the aggregation MSE in isolation at each communication round. For each round, the MSE minimization problem has been solved in \cite{Cao_PowerTWC}\footnote{Although the conventional channel inversion power control can achieve the unbiased aggregation, it is not the only way to achieve the unbiased aggregation and just a sufficient condition leading to unbiased aggregation. Moreover, as validated in \cite{Cao_PowerTWC}, the conventional MSE minimization scheme can achieve the minimum communication distortion in AirComp. Therefore, in this paper we only consider the conventional MSE minimization scheme as one benchmark, which always outperforms the generally sub-optimal channel inversion scheme.
%Thus, the channel inversion scheme can be replaced by the conventional MSE minimization scheme.
	}.
\end{itemize}
%{\color{blue}For ease of description, we define the cases without and with unbiased aggregation constraints as {\it Case I} and {\it Case II}, respectively.}

\subsection{Air-FEEL for Ridge Regression}
First, we consider the ridge regression with the sample-wise loss function $f({\bf w},{\bf x},\tau)=\frac{1}{2}\| {\bf x}^T{\bf w}-\tau\|^2+\rho R({\bf w})$ \footnote{
The loss function here consists of both the model loss and the regularization term, where the former captures the prediction error of the trained model over the samples, and the latter is added to avoid overfitting and  enhance the robustness \cite{LChen20TWC}.} and the regularization function $R({\bf w})=\|{\bf w}\|^2$ with the hyperparameter  $\rho=5\times 10^{-5}$.
Randoml,y generated synthetic dataset is used for model training and testing.
%, where another synthetic dataset with size of $200$ pairs (${\bf x}$, $y$), are for prediction. 
The generated data sample vector ${\bf x}\in\mathbb{R}^q$ with $q=10$ follow i.i.d. Gaussian distribution (i.e., $ x \sim\mathcal{N}(0,{\bf I})$) and the label $y$ is obtained as $\tau=x(2)+3x(5)+0.2z$, where $x(t)$ represents the $t$-th in vector ${\bf x}$ and $z$ denotes the observation noise with i.i.d. Gaussian distribution, i.e., $z\sim\mathcal{N}(0,1)$.
Unless stated otherwise, the data samples are evenly distributed among devices with identical size $ D =  D_k=1000, \forall k\in \mathcal K$ and $D_{\rm tot} = \sum\limits_{k\in \mathcal{K} } D_k = 10000.  $

% as well as the learning rate is set to be $\eta=0.05$.
Based on the above models, we can obtain the smoothness parameter $L$ and Polyak-{\L}ojasiewicz parameter $\delta$ as the largest and smallest eigenvalues of the data Gramian matrix ${\bf X}^T{\bf X}/D_{\rm tot}+10^{-4}{\bf I}$, in which ${\bf X}=[{\bf x}_1,\cdots,{\bf x}_{D_{\rm tot}}]^T$ is the data matrix. We use the  simple upper bounds $G^{(n)} =2WL $  \cite{DLiu2020Ar} with $W\geq \|\bf w\|$ as a bound on the norm $ \|\bf w\|$.
The optimal loss function $F^{\star}$ is computed using the optimal parameter vector $\bf w^{\star}$ to the learning problem \eqref{OptimalParameter}, where ${\bf w}^{\star}=({\bf X}^T{\bf X}+\rho{\bf I})^{-1}{\bf X}^T\boldsymbol{\tau}$ with $\boldsymbol{\tau}=[\tau_1,\cdots,\tau_{D_{\rm tot}}]^T$.
We set the initial parameter vector as an all-zero vector. 

\begin{figure*}[htbp] %\vspace{-0.05cm}
  \centering
  \subfigure[Optimality gap versus $N$ under diminishing learning rate.]
  {\label{DL_fig:OG_v_N}\includegraphics[width=8.6cm]{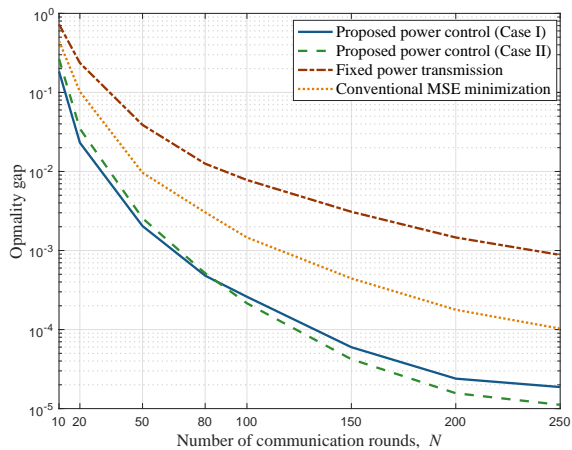}}
%   \vspace{-0.05in}
  \subfigure[Prediction error versus $N$ under diminishing learning rate.]
  {\label{DL_fig:PE_v_N}
\includegraphics[width=8.6cm]{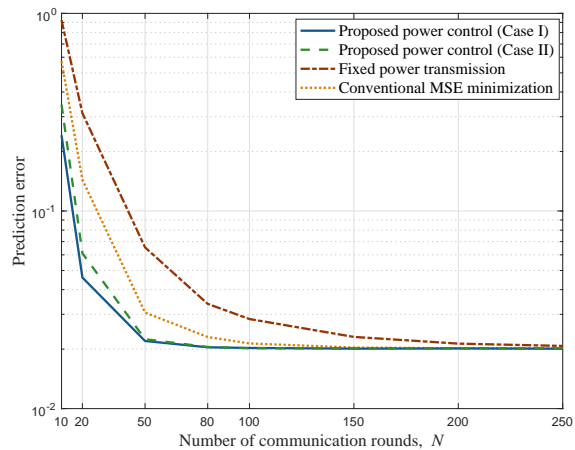}}
%    \vspace{-0.05in}
  \subfigure[Optimality gap versus $N$ under fixed learning rate.]
  {\label{fig:OG_v_N}\includegraphics[width=8.6cm]{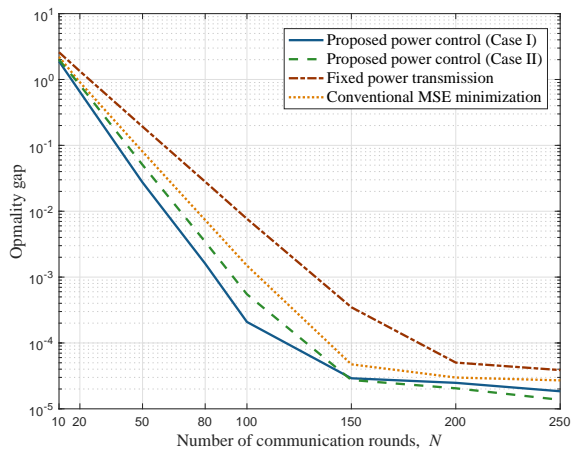}}
%   \vspace{-0.05in}
  \subfigure[Prediction error versus $N$ under fixed learning rate.]
  {\label{fig:PE_v_N}
\includegraphics[width=8.6cm]{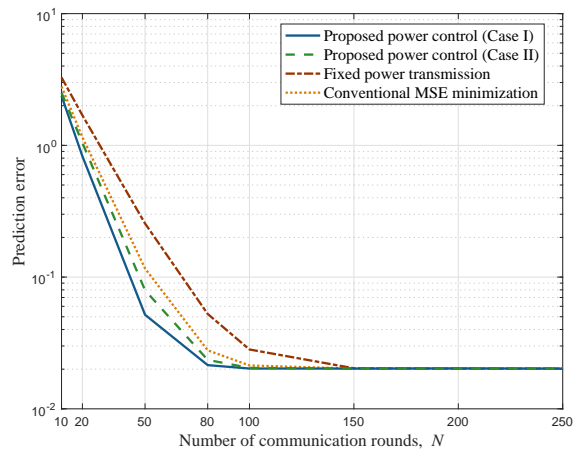}}
  \caption{Learning performance of Air-FEEL over number of communication rounds.}
  \label{Fig:Learning_v_N}
%\vspace{-0.4cm}
\end{figure*}

Fig.~\ref{Fig:Learning_v_N} shows the learning performance
(i.e., the optimality gap in Figs. \ref{DL_fig:OG_v_N} and \ref{fig:OG_v_N} and the prediction error in Figs. \ref{DL_fig:PE_v_N} and \ref{fig:PE_v_N})
 versus the number of communication rounds $N$, where the learning rates are set to be diminishing in Figs.~\ref{DL_fig:OG_v_N} and~\ref{DL_fig:PE_v_N} and those are set to be fixed in Figs.~\ref{fig:OG_v_N} and~\ref{fig:PE_v_N}.
First, it is observed that the proposed power control policies (with both biased and unbiased aggregation constraints) and the conventional MSE minimization design achieve faster convergence and lower optimality gap than the fixed power transmission. This shows the benefit of power control optimization in accelerating the learning convergence rate, via either directly minimizing the optimality gap or indirectly minimizing the MSE. 
Secondly, the proposed power control policies are observed to  significantly outperform the conventional MSE minimization design in reducing the optimality gap. This is due to the fact that the contributions of aggregation errors to the optimality gap are distinct at different communication rounds (see Remark~\ref{Remark_ContractionRegion}), which cannot be captured by the conventional MSE minimization design.
Furthermore, the proposed power control policy under Case II (with unbiased aggregation constraints) is observed to achieve a lower optimality gap than the proposed power control policy under Case I (without unbiased aggregation constraints) when $N>150$ under the fixed learning rate and $N>80$ under the  diminishing learning rates. This coincides with Remark~\ref{Remark_ContractionRegion} that the Air-FEEL algorithm converges to the optimal point with unbiased gradient aggregation.
%In addition, by comparing Fig. 3(a) (or  3(b)) versus Fig. 3(c) (or 3(d)), both the biased- and unbiased-aggregation schemes could achieve faster convergence in the diminishing learning rate case as depicted in Figs.~\ref{DL_fig:OG_v_N} and~\ref{DL_fig:PE_v_N}, than that in the fixed learning rate case as shown in Figs.~\ref{fig:OG_v_N} and~\ref{fig:PE_v_N}, which benefited from the adaptive learning rate.

\begin{figure*}[htbp]  %\vspace{-0.05cm}
  \centering
  \subfigure[Optimality gap versus $K$ under diminishing learning rate.]
  {\label{fig:FL_v_DL_K1}\includegraphics[width=8.6cm]{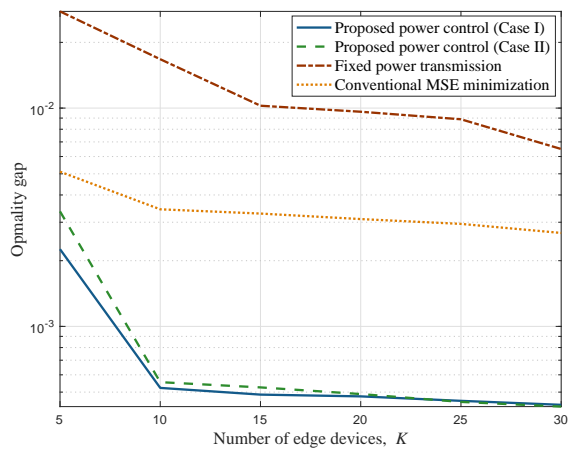}}
%  \vspace{0.08in}
  \subfigure[Prediction error versus $K$ under diminishing learning rate.]
  {\label{fig:FL_v_DL_K2}
\includegraphics[width=8.6cm]{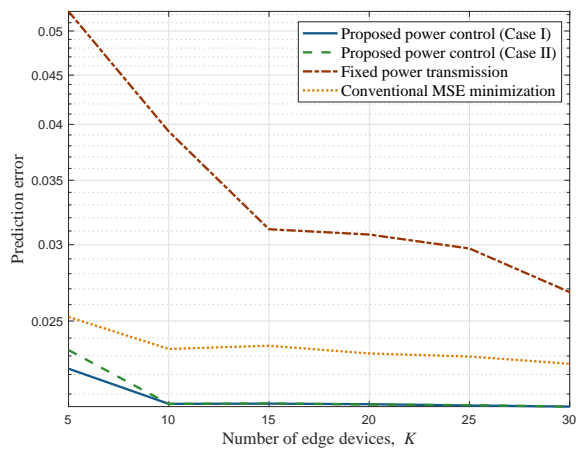}}
  \subfigure[Optimality gap versus $K$ under fixed learning rate.]
  {\label{fig:FL_v_K1}\includegraphics[width=8.6cm]{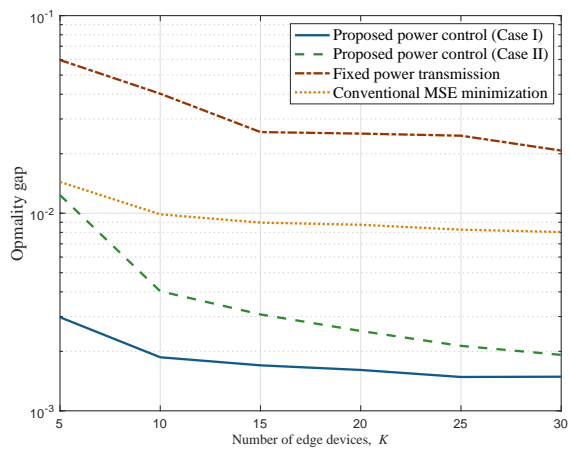}}
%  \vspace{0.08in}
  \subfigure[Prediction error versus $K$ under fixed learning rate.]
  {\label{fig:FL_v_K2}
\includegraphics[width=8.6cm]{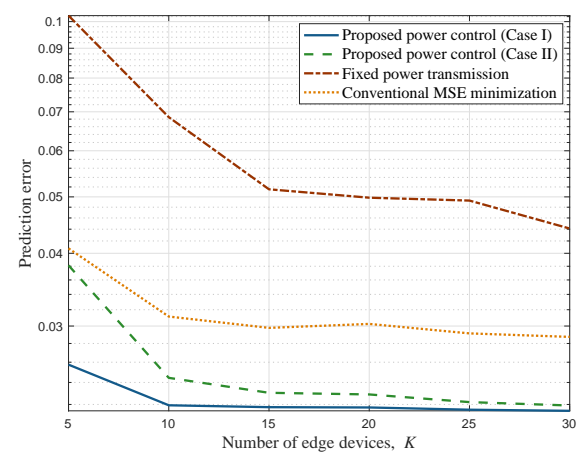}}
%   \vspace{-0.05in}
  \caption{Effect of number of devices on the learning performance of  Air-FEEL.}
  \label{Fig:FL_v_K}
%\vspace{-0.3cm}
\end{figure*}
Fig.~\ref{Fig:FL_v_K} shows the learning performance (i.e., the optimality gap in Figs. \ref{fig:FL_v_DL_K1} and \ref{fig:FL_v_K1} and the prediction error in Figs. \ref{fig:FL_v_DL_K2} and \ref{fig:FL_v_K2}) versus the number of devices $K$, where the learning rates are set to be diminishing in Figs.~\ref{fig:FL_v_DL_K1} and~\ref{fig:FL_v_DL_K2} and those are set to be fixed in Figs.~\ref{fig:FL_v_K1} and~\ref{fig:FL_v_K2}.
Firstly, it is observed that the optimality gap achieved by all schemes decreases as $K$ increases. This is because that the edge server can aggregate more data for averaging to improve the learning performance.
Secondly, the performance gaps between the proposed power control policies (under Cases I and II) versus the benchmark schemes are observed to decrease with $K$ increasing, which would be saturated in the large $K$ regime.
The performance gap validates the effectiveness on the proposed power control optimization in reducing the optimality gap.

\subsection{Air-FEEL for Handwritten Digit Recognition}\label{Sec_CNN}
\begin{figure*}[htbp] \vspace{-0.05cm}
  \centering
  \subfigure[Loss value versus $N$ under diminishing learning rate.]
  {\label{CNN_DL_fig:OG_v_N}\includegraphics[width=8.6cm]{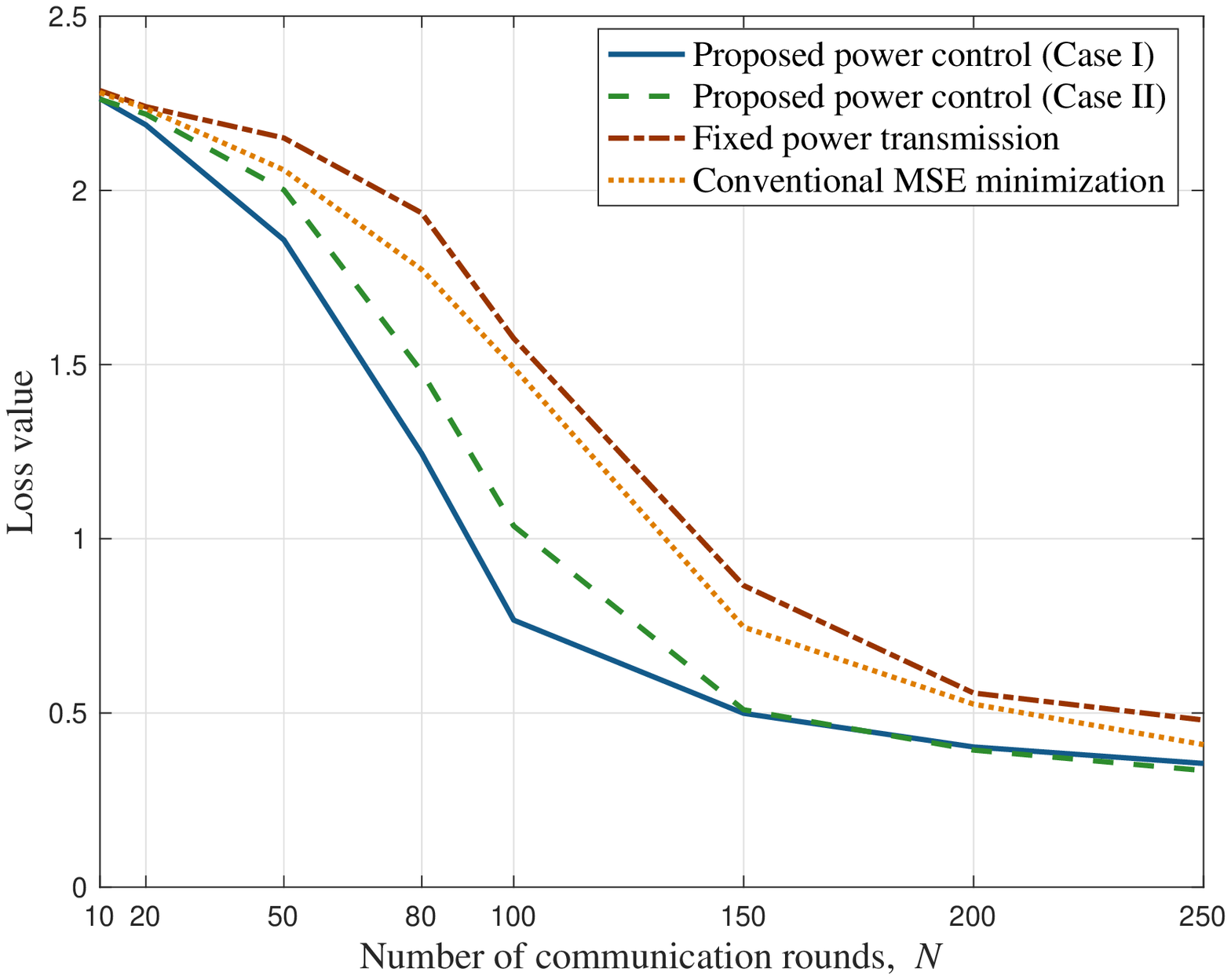}}
%   \vspace{-0.05in}
  \subfigure[Test accuracy versus $N$ under diminishing learning rate.]
  {\label{CNN_DL_fig:PE_v_N}
\includegraphics[width=8.6cm]{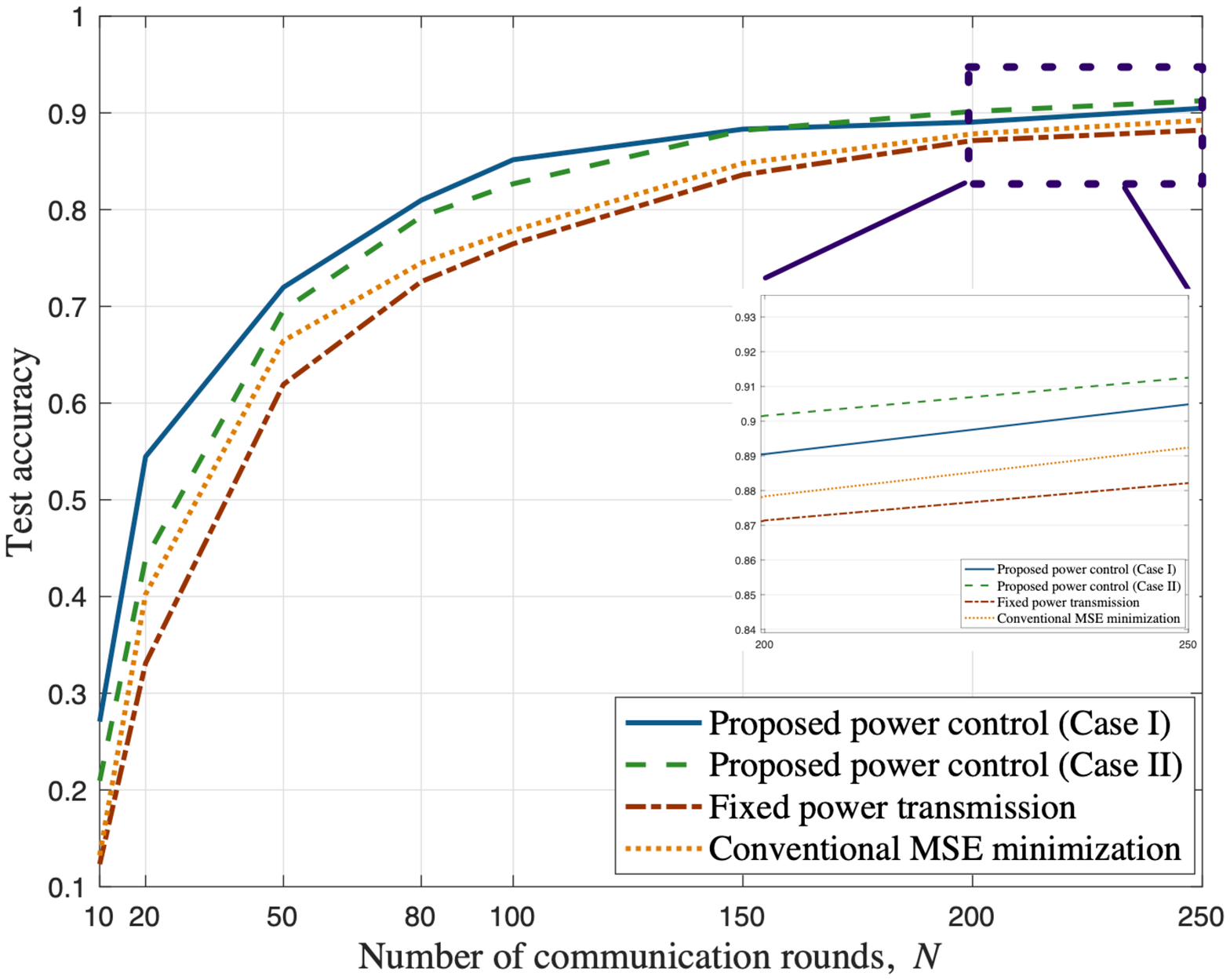}}
  \subfigure[Loss value versus $N$ under fixed learning rate.]
  {\label{CNN_fig:OG_v_N}\includegraphics[width=8.6cm]{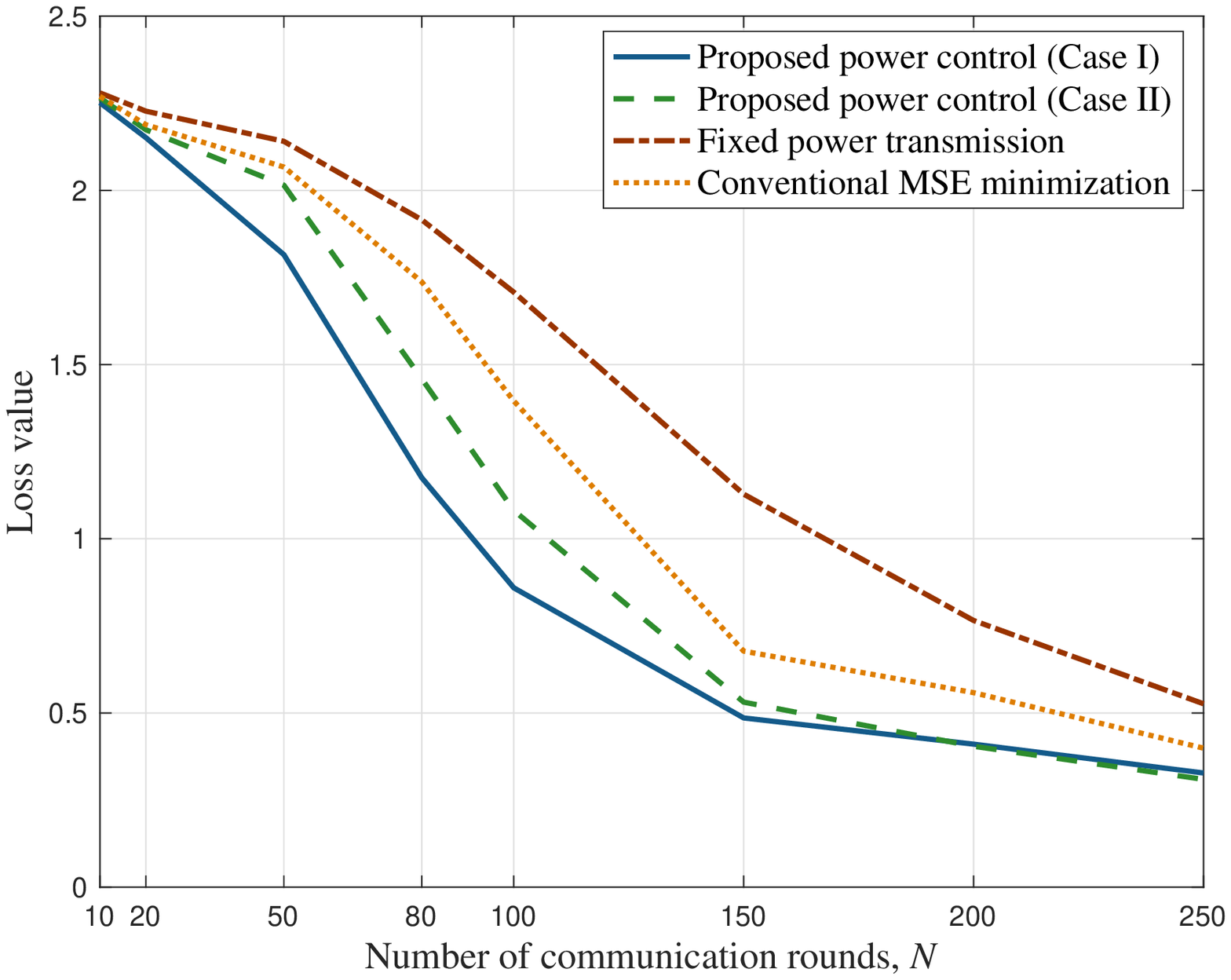}}
%   \vspace{-0.05in}
  \subfigure[Test accuracy versus $N$ under fixed learning rate.]
  {\label{CNN_fig:PE_v_N}
\includegraphics[width=8.6cm]{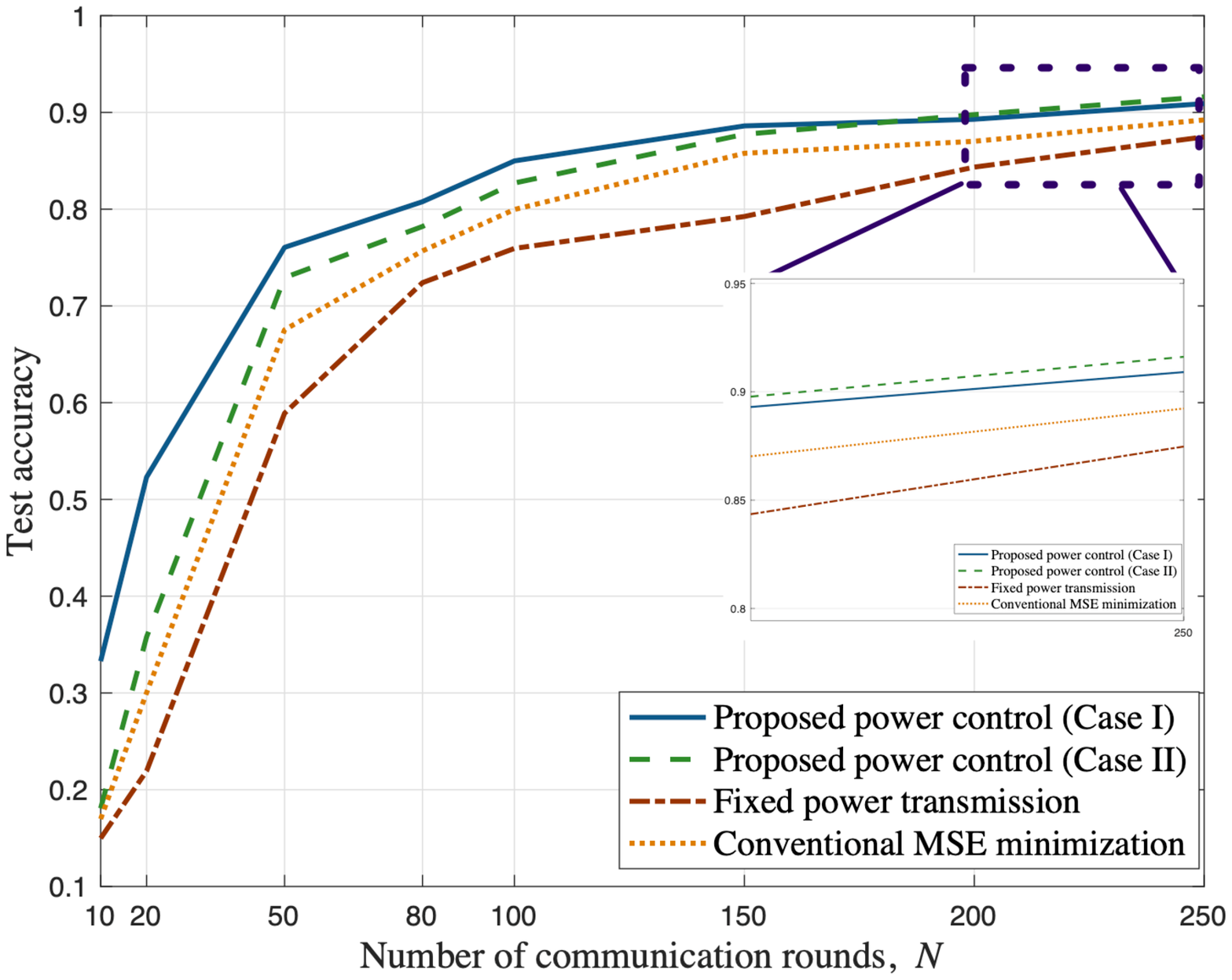}}
%    \vspace{-0.05in}
  \caption{Learning performance of Air-FEEL on MNIST dataset over number of communication rounds.}
  \label{CNN_Fig:N}
%\vspace{-0.4cm}
\end{figure*}

Next, we consider the learning task of handwritten digit recognition using the well-known MNIST datasets, which consists of 10 classes of black-and-white digits ranging from ``0" to ``9". %each image has 784 pixels and a label. 
We implement  a 6-layer CNN as the classifier model, which  consists of two $5\times 5$ convolution layers with ReLU activation (the first with 32 channels, the second with 64), each followed by a $2 \times 2$ max pooling; a fully connected layer with 512 units and ReLU activation; and a final softmax output layer ($582,026$ parameter in total). The local batch size at each edge device is set to be $m_b=512$.
Notice that  Assumptions \ref{Assump_Smooth} and \ref{Assump_PL} may not hold in this case, but our proposed power control policies still work well as will be shown shortly.

Fig.~\ref{CNN_Fig:N} shows the learning performance versus the varying number of communication rounds $N$, where the learning rates are set to be diminishing in Figs.~\ref{CNN_DL_fig:OG_v_N} and~\ref{CNN_DL_fig:PE_v_N} and those are set to be fixed with $\eta=0.01$ in Figs.~\ref{CNN_fig:OG_v_N} and~\ref{CNN_fig:PE_v_N}.
%where the learning rate at Figs.~\ref{CNN_fig:OG_v_N} and~\ref{CNN_fig:PE_v_N} is set to be diminishing and that in Figs.~\ref{CNN_fig:OG_v_N} and~\ref{CNN_fig:PE_v_N} is fixed to be $0.01$.
First, it is observed that the proposed power control policies achieve lower loss function values and higher test accuracy than both the fixed-power-transmission and conventional-MSE-minimization schemes. 
Furthermore, the power control policy under Case II is observed to outperform that under Case I when $N>200$ with the  fixed learning rate and $N>150$ with the diminishing learning rates. These observations are generally consistent with those in Fig. \ref{Fig:Learning_v_N} with the ridge regression model.

\section{Conclusion}%\vspace{-0.2cm}

In this paper, we exploited the transmission power control as a new design degree of freedom to optimize the learning performance of Air-FEEL. To this end, we first analyzed the convergence behavior of the FEEL algorithm (in terms of the optimality gap) and characterized the impact of aggregation errors, w.r.t. its bias and MSE, at different communication rounds. 
It was observed that in the case with unbiased aggregation estimates, the FEEL algorithm would converge exactly to the optimal point with mild conditions; and otherwise, it would converge with an error floor. 
Next, we proposed to directly minimize the derived optimality gaps by optimizing the power control, for which the optimal solutions are obtained to follow regularized channel inversion structures. 
Finally, experimental results demonstrated that the proposed power control policies achieve significantly lower optimality gap in Air-FEEL, as compared with benchmark schemes with fixed power transmission and conventional MSE minimization.
We expect that this initial work can provide useful insights on exploiting the power control for enhancing the Air-FEEL performance.

Nevertheless, due to the space limitation, there are still  a lot of interesting issues that are not addressed  in this paper but worth investigation. 
 In the following, we introduce several of them to motivate  future work. 
\begin{itemize}
	\item  One interesting direction is to explore the large-scale Air-FEEL over multi-cell networks to accommodate massive edge devices with more data. In this case, a hierarchical framework would be established to explore the distributed computation and communication capacity for performance improvement, and therein the cooperative interference management will be a new issue to be addressed.
    \item Another interesting direction is to consider the FedAvg scenario, where the local updates at each edge devices are implemented multiple times, and the local models are aggregated over the air instead of gradients. The analysis and design principles in this paper are generally extendable to this scenario by taking into account the following two new technical challenges. First, as multiple local updates are implemented at edge devices, the introduced gradient/model errors due to aggregation in this paper are not applicable for FedAvg. Therefore, new  approaches for analyzing the gradient/model errors in FedAvg should be considered for its convergence analysis. Next, while the learning rate plays a dual role of denoising factor for over-the-air gradient aggregation in FedSGD, in FedAvg we need a dedicated denoising factor to  suppress the AirComp signal misalignment error for model aggregation. As a result,  the denoising factors would become new design variables for joint optimization, thus making the problem more complicated.
\item Furthermore, it is also an interesting problem to fairly and quantitatively compare the performance of Air-FEEL versus the traditional digital FEEL (e.g., \cite{Chen20FL,Mo2020aa,Tran19_Infocom}), in terms of the training latency and training accuracy.
 General speaking, the proposed Air-FEEL can achieve significant per-round communication latency reduction over the digital FEEL, but at a cost of the newly introduced aggregation errors and the increased number of communication rounds needed for convergence. To deal with such a tradeoff, we need to explore the power control policy for training latency minimization while ensuring a given maximum optimality gap requirement for both Air-FEEL and digital FEEL. How to determine proper approximate optimality gaps for both Air-FEEL and digital FEEL to enable their fair training performance comparison is challenging in practice.  
\item Moreover, the investigation of Air-FEEL with the non-i.i.d. data is also an interesting future research direction. 
The non-i.i.d. nature of data may degrade the training performance of AirFEEL, and also make the convergence analysis more challenging. In this case, how to capture the effect of the non-i.i.d. degrees of edge devices on the learning performance and accordingly optimize the transmission power control over them is a difficult task.
\end{itemize}

% One interesting direction is to explore the large-scale Air-FEEL over multi-cell networks, where cooperative interference management will be a new issue to be tackled. 
%\cite{Kang20,SWang19_FedAvg}

%This work can also be extended to the FedAvg scenario with the training model aggregating, which is different from the FedSGD case.
%  First, the learning rate can be reused to play the role of denoising factors for {\it over-the-air computation} (AirComp) in the FedSGD case, while the dedicated denoising factors should be used for suppressing the AirComp signal-misalignment error in Air-FedAvg. 
%  Besides, the gradient aggregation in FedSGD can provide guaranteed training convergence, but at a cost of heavy communication costs caused by the frequent gradient aggregation in each every single local gradient update; thus,  the hyper-parameters (i.e., the number of local update epochs) provides new design parameters in FedAvg to accelerate the learning convergence.

\appendix

\subsection{Proof of Theorem~\ref{Theo_OG_FixedRate}}\label{Theo_OG_FixedRate_Proof}
The proof follows by relating the norm of the gradient to the expected improvement made at each communication round.
Recall that $\!{\bf w}^{(n+1)}\!={\bf w}^{(n)}-\!\eta\cdot\! \left(\bar{\bf g}^{(n)}+{\bm \varepsilon}^{(n)}\right)$, and thus it follows that
\begin{align}
	&F\left({\bf w}^{(n+1)}\right)- F\left({\bf w}^{(n)}\right)\notag\\
	&\leq  \! \left(\!\nabla F\left({\bf w}^{(n)}\!\right)\!\!\right)^T\! ({\bf w}^{(n+1)}\!-\! {\bf w}^{(n)}) \!+\! \sum_{i=1}^{q}\!\frac{L_i}{2}\left({{ w}_i^{(n+1)}\!-\!{w}^{(n)}_i}\!\right)^2\!\notag\\
	&= \nabla F\left({\bf w}^{(n)}\right)^T ({\bf w}^{(n)}-\eta\cdot \left(\bar{\bf g}^{(n)}+{\bm \varepsilon}^{(n)}\right)- {\bf w}^{(n)})\notag\\ 
	&~~~+ \frac{L}{2}\left\|{\bf w}^{(n)}-\eta\cdot \left(\bar{\bf g}^{(n)}+{\bm \varepsilon}^{(n)}\right)- {\bf w}^{(n)}\right\|^2\notag
	\end{align}
	\begin{align}
	&=-\eta \nabla F\left(\!{\bf w}^{(n)}\!\right)^T\! \!\left(\bar{\bf g}^{(n)}\!+\!{\bm \varepsilon}^{(n)}\right)\!\!+ \!\frac{L\eta^2}{2}\!\left\|\bar{\bf g}^{(n)}+{\bm \varepsilon}^{(n)}\right\|^2\!,\!\!\label{Proof_Smoth1}
\end{align}
where the above inequality follows Assumption~\ref{Assump_Smooth} and $L\triangleq \|\bf L\|_{\infty}$.
By taking expectation at both sides of \eqref{Proof_Smoth1}, we have
\begin{align}
&\mathbb{E}\left[F\left({\bf w}^{(n+1)}\right)-F\left({\bf w}^{(n)}\right) \right]\notag\\
	&\!\le\! - \eta \nabla F\left(\!{\bf w}^{(n)}\!\right)^T\!\! \mathbb{E}\left[\bar{\bf g}^{(n)}\!+\!{\bm \varepsilon}^{(n)}\right]+ \!\frac{L\eta^2}{2}\mathbb{E}\left[\left\|\bar{\bf g}^{(n)}\!+\!{\bm \varepsilon}^{(n)}\right\|_2^2\right]\notag\\
&\!=\!-\!\eta \left\|\nabla\!F\!\left(\!{\bf w}^{(n)}\!\right)\!\right\|^2\!\!\!- \!\eta\nabla F\left({\bf w}^{(n)}\!\right)^T\!\!\mathbb{E}\!\left[\!{\bm \varepsilon}^{(n)}\!\right]\!+\!\frac{L\eta^2}{2}\mathbb{E}\left[\left\|\bar{\bf g}^{(n)}\right\|^2\!\right]\notag\\
&~~+L\eta^2\nabla F\left({\bf w}^{(n)}\right)^T \mathbb{E}\left[{\bm \varepsilon}^{(n)}\right]+\frac{L\eta^2}{2}\mathbb{E}\left[\left\|{\bm \varepsilon}^{(n)}\right\|^2\right]\notag\\
&\leq-\eta \left\|\nabla F\left({\bf w}^{(n)}\right)\right\|^2- \eta\nabla F\left({\bf w}^{(n)}\right)^T\mathbb{E}\left[{\bm \varepsilon}^{(n)}\right]\notag\\
&~~+\frac{L\eta^2}{2}\left(\frac{\left\|\nabla F\left({\bf w}^{(n)}\right)\right\|_2^2}{K}+\frac{\left\|{\bm\sigma}\right\|_2^2}{m_bK^2}\right)\notag\\
&~~+\frac{L\eta^2}{2}\mathbb{E}\left[\left\|{\bm \varepsilon}^{(n)}\right\|^2\right]+L\eta^2\nabla F\left({\bf w}^{(n)}\right)^T \mathbb{E}\left[{\bm \varepsilon}^{(n)}\right],\label{Proof1_eq1}
\end{align}
\!where the denominator $m_b$ in \eqref{Proof1_eq1} is induced from Assumption \ref{Assum_VarianceBound} and Equation \eqref{sys_LocalGradient}. 
By applying the inequality of arithmetic and geometric means, i.e., $- {\bm a_1}^T{\bm a_2}\leq \frac{\|{\bm a_1}\|^2 }{2}+\frac{\|{\bm a_2}\|^2 }{2}$,  it then follows that 
\begin{align*}
&\mathbb{E}\left[F\left({\bf w}^{(n+1)}\right)-F\left({\bf w}^{(n)}\right) \right]\notag\\
&\leq-\eta \left\|\nabla F\left({\bf w}^{(n)}\right)\right\|^2+\frac{\eta^2}{2}\left\|\nabla F\left({\bf w}^{(n)}\right)\right\|^2+\frac{\left\|\mathbb{E}\left[{\bm \varepsilon}^{(n)}\right]\right\|^2}{2}\notag\\
&~~+\frac{L\eta^2}{2}\mathbb{E}\left[\left\|{\bm \varepsilon}^{(n)}\right\|^2\right]+L\eta^2\nabla F\left({\bf w}^{(n)}\right)^T \mathbb{E}\left[{\bm \varepsilon}^{(n)}\right]\notag\\
&~~+\frac{L\eta^2}{2}\left(\left\|\nabla F\left({\bf w}^{(n)}\right)\right\|_2^2+\frac{\left\|{\bm\sigma}\right\|_2^2}{m_bK^2}\right).%\label{Proof1_ineq1}.
\end{align*}
%
%while the inequalities \eqref{Proof1_ineq1} and \eqref{Proof1_ineq2} follows the inequality of arithmetic and geometric means, i.e. 
%$- {\bm a_1}^T{\bm a_2}\leq \frac{\|{\bm a_1}\|^2 }{2}+\frac{\|{\bm a_2}\|^2 }{2}$ and 
Under the Cauchy-Schwarz's inequality, i.e., $ {\bm b_1}^T{\bm b_2}\leq \|{\bm b_1}\|\|\bm b_2\|\leq\frac{\|{\bm b_1}\|^2 }{2}+\frac{\|{\bm b_2}\|^2 }{2}$, we further have 
\begin{align}
&\mathbb{E}\left[F\left({\bf w}^{(n+1)}\right)-F\left({\bf w}^{(n)}\right) \right]\notag\\
&\leq-\eta \left\|\nabla F\left({\bf w}^{(n)}\right)\right\|^2+\frac{\eta^2\left\|\nabla F\left({\bf w}^{(n)}\right)\right\|^2}{2}+\frac{\left\|\mathbb{E}\left[{\bm \varepsilon}^{(n)}\right]\right\|^2}{2}\notag\\
&~~+\frac{L\eta^2}{2}\mathbb{E}\left[\left\|{\bm \varepsilon}^{(n)}\right\|^2\right]+\frac{\eta^2L^2\left\|\mathbb{E}\left[{\bm \varepsilon}^{(n)}\right]\right\|^2}{2}\notag
\end{align}
\begin{align}
&~~+\frac{L\eta^2}{2}\left(\left\|\nabla F\left({\bf w}^{(n)}\right)\right\|_2^2+\frac{\left\|{\bm\sigma}\right\|_2^2}{m_bK^2}\right)+\frac{\eta^2\left\|\nabla F\left({\bf w}^{(n)}\right)\right\|^2}{2}\notag\\
&=-\eta\left(1-\eta-\frac{\eta L}{2} \right) \left\|\nabla F\left({\bf w}^{(n)}\right)\right\|^2+\frac{\eta^2L\left\|{\bm\sigma}\right\|_2^2}{2m_bK^2}\notag\\
&~~+\frac{\left(1+\eta^2L^2\right)\left\|\mathbb{E}\left[{\bm \varepsilon}^{(n)}\right]\right\|^2}{2}+\frac{\eta^2L}{2}\mathbb{E}\left[\left\|{\bm \varepsilon}^{(n)}\right\|^2\right]\notag\\
&\leq-\frac{\eta}{2}\left\|\nabla F\left({\bf w}^{(n)}\right)\right\|^2+\frac{\left(1+\eta^2L^2\right)\left\|\mathbb{E}\left[{\bm \varepsilon}^{(n)}\right]\right\|^2}{2}\notag\\
&~~+\frac{\eta^2L\left\|{\bm\sigma}\right\|_2^2}{2m_bK^2}+\frac{\eta^2L}{2}\mathbb{E}\left[\left\|{\bm \varepsilon}^{(n)}\right\|^2\right]\label{Proof1_ineq3},
\end{align}
where the inequality in \eqref{Proof1_ineq3} follows that  $\eta \leq\frac{2}{2+L}$.
%where the denominator $m_b$ in \eqref{Proof1_eq1} is induced from Assumption \ref{Assum_VarianceBound} and Equation \eqref{sys_LocalGradient}, %which is because that the local gradient estimate is computed on the local loss function over a mini-batch of size $m_b$, thus the resultant gradient variance is reduced from $\left\|{\bm\sigma}\right\|_2^2$ to $\frac{\left\|{\bm\sigma}\right\|_2^2}{m_b}$;
% the inequality \eqref{Proof1_ineq1} holds by relaxing the $\eta^2$ in the last term in equality \eqref{Proof1_eq1};
%while the inequalities \eqref{Proof1_ineq1} and \eqref{Proof1_ineq2} follows the inequality of arithmetic and geometric means, i.e. 
%$- {\bm a_1}^T{\bm a_2}\leq \frac{\|{\bm a_1}\|^2 }{2}+\frac{\|{\bm a_2}\|^2 }{2}$ and Cauchy-Schwarz's inequality, i.e., $ {\bm b_1}^T{\bm b_2}\leq \|{\bm b_1}\|\|\bm b_2\|\leq\frac{\|{\bm b_1}\|^2 }{2}+\frac{\|{\bm b_2}\|^2 }{2}$, respectively;
Next, by applying Assumption \ref{Assump_PL}, it hence follows that
\begin{align*}
&\mathbb{E}\left[F\left({\bf w}^{(n+1)}\right)\right]-F^{\star}\leq \left(1-\delta\eta\right)\left(\mathbb{E}\left[F\left({\bf w}^{(n)}\right)\right]-F^{\star}\right) \notag\\
&~+\frac{\left(\!1+\eta^2L^2\right)\left\|\mathbb{E}\left[{\bm \varepsilon}^{(n)}\right]\right\|^2}{2}+\frac{\eta^2L\left\|{\bm\sigma}\right\|_2^2}{2m_bK^2}+\!\frac{\eta^2L}{2}\mathbb{E}\left[\left\|{\bm \varepsilon}^{(n)}\right\|^2\right].
\end{align*}
Through some further algebraic manipulation over the above inequality based on $m_b = N$ and $C=1-\delta\eta$, we have \eqref{Gap_Iterate}.
This thus completes the proof.

\subsection{Proof of Proposition~\ref{lemma_Biased_Power}}\label{Proof_Lemma_Biased}

Notice that problem (P1.1) is convex, and thus strong duality holds between problem (P1.1) and its Lagrange dual problem \cite{cvx}. Hence, we leverage the Lagrange duality method to optimally solve problem (P1.1).
Let $\varphi_k\ge 0, \forall k\in\cal K$ denote the dual variable associated with the $k$-th constraint in \eqref{Biased_AvePower}, respectively. 
The partial Lagrangian of problem (P1.1) is thus given by
\begin{align*}
\mathcal{L}_1\left(\left\{\hat{p}_k^{(n)}\right\}\right)=& \sum_{n=1}^{N}J^{(n)}A^{(n)}\left(\sum\limits_{k\in\mathcal K}h_k^{(n)}\hat{p}_k^{(n)}-K\right)^2\notag\\
&+\sum_{n=1}^{N}J^{(n)}B^{(n)}\sum\limits_{k\in\mathcal K}\left(h_k^{(n)}\hat{p}_k^{(n)}-1\right)^2\\
&+\sum\limits_{k\in\mathcal{K}} \varphi_k\left(\frac{1}{N}\sum \limits_{n\in\mathcal{N}}\left(\hat{p}_{k}^{(n)}\right)^2\hat{G}^{(n)} - P^{\rm ave}_k\right).
\end{align*}
Then the dual function is
\begin{align}\label{Biased_DualFucntion}
W_1(\{\varphi_k\})=\min_{\{0\leq q_k^{(n)}\leq P^{\rm max}_{k,n}\}} ~&\mathcal{L}_1\left(\left\{\hat{p}_k^{(n)}\right\}\right).
%{\rm s.t.}~~&\eqref{Biased_MaxPower}.\notag
\end{align}
The dual problem of problem (P1) is given as
\begin{align}
\mathbf{D1:} \min_{\{\varphi_k\ge 0\}} ~&W_2(\{\varphi_k\}).
\end{align}
Due to the strong duality between problems (P1.1) and (D1), we solve problem (P1.1) by equivalently solving its dual problem (D1). Let $\{\hat{p}_k^{(n)\rm opt}\}$ denote the optimal primal solution to problem (P1.1), and $\{\varphi_k^{\rm opt}\}$ the optimal dual solution to problem (D1). We first evaluate the dual function $W_1(\{\varphi_k\})$ under any given feasible $\{\varphi_k\}$, and then obtain the optimal dual variables $\{\varphi_k^{\rm opt}\}$ to maximize $W_1(\{\varphi_k\}\})$.

First, we obtain $W_1(\{\varphi_k\}\})$ by solving problem \eqref{Biased_DualFucntion} under any given feasible $\{\varphi_k\}$, which can be decomposed into a series of subproblems each for one communication round $n$ as follows.
\begin{align}\label{Biased_power_n}
\!\!\!\!\min_{\{q_k^{(n)}\}} ~~&J^{(n)}A^{(n)}\left(\sum\limits_{k\in\mathcal K}h_k^{(n)}\hat{p}_k^{(n)}-K\right)^2\!\!\!\!+\frac{\varphi_k\hat{G}^{(n)}\left(\hat{p}_{k}^{(n)}\right)^2}{N}\notag\\
&~+J^{(n)}B^{(n)}\sum\limits_{k\in\mathcal K}\left(h_k^{(n)}\hat{p}_k^{(n)}\!-\!1\!\right)^2 \\
{\rm s.t.}~&0\leq q_k^{(n)}\leq P^{\rm max}_{k,n}, \forall k\in\mathcal{K}.
\end{align}
By taking its first-order derivation w.r.t. each $\hat{p}_k^{(n)}$, we have the following lemma.
\begin{lemma}\label{lemma_Biased_power_n}\emph{The optimal solution to problem \eqref{Biased_power_n} denoted by $q_k^{(n)\star}, \forall k\in\mathcal{K},~n\in\mathcal{N}$ is given as
\begin{align}
\hat{p}_k^{(n)\star}=\min\left[\frac{B^{(n)}+A^{(n)}K }{M_k^{(n)}+ A^{(n)}M_k^{(n)}\sum\limits_{i\in\mathcal{K}} \frac{h_i^{(n)}}{M_i^{(n)}} } ,P^{\rm max}_{k,n}\right],
\end{align}
where $ M_k^{(n)}\triangleq B^{(n)} h_k^{(n)}+\frac{ \varphi_k\hat{G}^{(n)}}{NJ^{(n)}h_k^{(n)}} , \forall k\in\mathcal{K},~n\in\mathcal{N} $.
}
\end{lemma}
With Lemma~\ref{lemma_Biased_power_n}, problem
\eqref{Biased_DualFucntion} is solved under any given $\{\varphi_k\}$, and the dual function $W_1(\{\varphi_k\}\})$ is accordingly obtained. It remains to find the optimal $\{\varphi_k\ge 0\}$.
Since the dual function $W_1(\{\varphi_k\})$ is concave but non-differentiable in general, one can use subgradient based methods such as the ellipsoid method \cite{Gradient} to obtain the optimal dual variables. Note that
for the objective function in \eqref{Biased_DualFucntion}, the subgradient w.r.t. $\varphi_k, \forall k$, is $\frac{1}{N}\sum \limits_{n\in\mathcal{N}}\left(\hat{p}_{k}^{(n)\star}\right)^2\hat{G}^{(n)} - P^{\rm ave}_k$. By replacing $\{\varphi_k\}$ in Lemma~ \ref{lemma_Biased_power_n} with the obtained optimal dual variables $\{\varphi_k^{\rm opt}\}$, the optimal solution to problem (P1.1) is accordingly obtained as shown in Proposition~\ref{lemma_Biased_Power}. This thus completes the proof.

\subsection{Proof of Proposition~\ref{theorem_Unbiased_gamma}}\label{Proof_theorem_Unbiased_gamma}

Notice that problem (P2.1) is convex and satisfies the Slater's condition, and thus strong duality holds between problem (P2.1) and its Lagrange dual problem \cite{cvx}. Therefore, we apply the Lagrange duality method to optimally solve problem (P2.1).
Let $\mu_n$ and $\lambda_k\ge 0$ denote the dual variable associated with  the $n$-th constraints in \eqref{Unbiased_alignment} and the $k$-th constraint in \eqref{Unbiased_AvePower}, respectively. Then the partial Lagrangian of problem (P2.1) is
\begin{align*}
\mathcal{L}\left(\{q_k^{(n)}\}\right)=&\sum_{n=1}^{N}J^{(n)}B^{(n)}\sum\limits_{k\in\mathcal K}\left(h_k^{(n)}q_k^{(n)}-1\right)^2\notag\\
&+\sum\limits_{k\in\mathcal{K}} \lambda_k\left(\frac{1}{N}\sum \limits_{n\in\mathcal{N}}\left(q_{k}^{(n)}\right)^2\hat{G}^{(n)} - P^{\rm ave}_k\right)\notag\\
&+\sum\limits_{n\in\mathcal{N}}\mu_n\left(\sum\limits_{k\in\mathcal K}h_k^{(n)}q_{k}^{(n)}-K\right).
\end{align*}
Then the dual function is
\begin{align}\label{Unbiased_DualFucntion}
W_2(\{\lambda_k\},\{\mu_n\})=\min_{\{q_k^{(n)}\ge 0\}} ~&\mathcal{L}\left(\{q_k^{(n)}\}\right)\\
{\rm s.t.}~~&\eqref{Unbiased_MaxPower}.\notag
\end{align}
Accordingly, the dual problem of problem (P1) is given as
\begin{align}
\mathbf{D2:} \min_{\{\lambda_k\ge 0\},\{\mu_n\}} ~&W_2(\{\lambda_k\},\{\mu_n\}).
\end{align}
Due to the fact that the strong duality holds between problems (P2.1) and (D2), we can solve problem (P2.1) by equivalently solving its dual problem (D2). For notational convenience, let $\{q_k^{(n)\rm opt}\}$ denote the optimal primal solution to problem (P2.1), and $\{\lambda_k^{\rm opt}\}$ and $\{\mu_n^{\rm opt}\}$ denote the optimal dual solution to problem (D2). In the following, we first evaluate the dual function $W_2(\{\lambda_k\},\{\mu_n\})$ under any given feasible $\{\lambda_k\}$ and $\{\mu_n\}$, and then obtain the optimal dual variables $\{\lambda_k^{\rm opt}\}$ and $\{\mu_n^{\rm opt}\}$ to maximize $W_2(\{\lambda_k\},\{\mu_n\})$.

First, we obtain $W_2(\{\lambda_k\},\{\mu_n\})$ by solving problem \eqref{Unbiased_DualFucntion} under any given feasible $\{\lambda_k\}$ and $\{\mu_n\}$, which can be decomposed into a sequence of subproblems each for optimizing the power scaling factor in edge device $k$ at one communication round $n$, i.e., 
\begin{align}\label{Unbiased_gammaPro}
\!\!\!\!\!\min_{q_k^{(n)}} ~&J^{(n)}\!B^{(n)}\!\left(\!h_k^{(n)}q_k^{(n)}\!\!-\!1\!\right)^2\!\!\!+\!\frac{\lambda_k\hat{G}^{(n)}}{N}\!\left(\!q_{k}^{(n)}\!\right)^2\!\!+\!\mu_nh_k^{(n)}\!q_{k}^{(n)}\!\!\!\\
\!\!\!\!\!\!\!{\rm s.t.}~&0\leq q_k^{(n)}\leq P^{\rm max}_{k,n}\notag
\end{align}
Via the first-order derivation of the objective function in \eqref{Unbiased_gammaPro}, we have the following lemma.

\begin{lemma}\label{lemma_Unbiased_gammaPro}\emph{The optimal solution to problem \eqref{Unbiased_gammaPro} denoted by $q_k^{(n)\star}, \forall k\in\mathcal{K},~n\in\mathcal{N}$ is given as
\begin{align}
q_k^{(n)\star}=\min\left[\frac{\left(h_k^{(n)}-\frac{\mu_nh_k^{(n)}}{2J^{(n)}B^{(n)}} \right)^+}{ (h_k^{(n)})^2+\frac{ 2\lambda_k\hat{G}^{(n)}}{NJ^{(n)}B^{(n)}}},P^{\rm max}_{k,n}\right].
\end{align}
}
\end{lemma}
Therefore, with Lemma \ref{lemma_Unbiased_gammaPro},  problem \eqref{Unbiased_DualFucntion} is solved, and the dual function $W_2(\{\lambda_k\},\{\mu_n\})$ is accordingly obtained. It remains to find the optimal $\{\lambda_k\ge 0\}$ and $\{\mu_n\}$.
Since the dual function $W_2(\{\lambda_k\},\{\mu_n\})$ is concave but non-differentiable in general, we can use the ellipsoid method \cite{Gradient},	 to obtain the optimal dual variables. Note that for the objective function in \eqref{Unbiased_DualFucntion}, the subgradient w.r.t. $\lambda_k, \forall k$, is $\frac{1}{N}\sum \limits_{n\in\mathcal{N}}\left(q_{k}^{(n)\star}\right)^2\hat{G}^{(n)} - P^{\rm ave}_k$, while that for dual variables $\mu_n, \forall n\in\cal N$, is $\sum\limits_{k\in\mathcal K}h_k^{(n)}q_{k}^{(n)\star}-K$. By replacing $\{\lambda_k\}$ and $\{\mu_n\}$ in Lemma~ \ref{lemma_Unbiased_gammaPro} with the obtained optimal dual variables $\{\lambda_k^{\rm opt}\}$ and $\{\mu_n^{\rm opt}\}$, the optimal solution to problem (P2.1) is accordingly obtained as shown in Lemma~\ref{theorem_Unbiased_gamma}. This thus completes the proof.

\bibliography{AirCompforFL}
\bibliographystyle{IEEEtran}

\end{document}